\documentclass[aps,prb,twocolumn,floatfix,citeautoscript,hyperref=pdftex]{revtex4-2}
\usepackage{graphicx}
\usepackage{dcolumn}
\usepackage{bm}
\usepackage{color}
\usepackage{amsmath}
\usepackage{amssymb}
\usepackage{multirow}
\newcommand {\beq} {\begin{equation}}
\newcommand {\eeq} {\end{equation}}
\newcommand {\bqa} {\begin{eqnarray}}
\newcommand {\eqa} {\end{eqnarray}}

\usepackage[colorlinks=true]{hyperref}

\begin{document}

\title{Constraints on superconducting pairing in altermagnets}

\author{Debmalya Chakraborty}
\thanks{Present address: Max Planck Institute for the Physics of Complex Systems, N\"othnitzer Stra\ss e 38, 01187, Dresden, Germany.}
\affiliation{Department of Physics and Astronomy, Uppsala University, Box 516, S-751 20 Uppsala, Sweden}

\author{Annica M. Black-Schaffer}
\affiliation{Department of Physics and Astronomy, Uppsala University, Box 516, S-751 20 Uppsala, Sweden}

\begin{abstract}

Superconductivity in the recently discovered altermagnetic materials hosts large prospects for both fundamental physics and technological applications. In this work we show that a characteristic spin-sublattice locking in altermagnets puts severe constraints on possible superconducting pairing. In particular, we uncover that the most common form of superconductivity, uniform  $s$-wave spin-singlet pairing is not possible to achieve in altermagnets. Considering an effective model for a $d_{x^2-y^2}$-wave altermagnet on a square lattice, we instead find that the most likely forms of spin-singlet pairing have  $d_{x^2-y^2}$- or extended $s$-wave symmetry. We also find that the simplest form of equal-spin-triplet $p$-wave pairing is not allowed, but it can only exist as a mixed-spin-triplet $p$-wave state.
We verify these constraints on pairing within an interaction-induced model of altermagnetism, where we also establish their validity for finite-momentum pairing. Additionally we discuss the possible pairing symmetries for odd-frequency superconducting pairing. Due to the generality of our results, they are applicable to both intrinsic superconductivity and proximity-induced superconductivity in altermagnet-superconductor hybrid junctions.
\end{abstract}

\maketitle

\section{Introduction}\label{sec:intro}

Magnetism and superconductivity are the two most celebrated quantum phases of matter and have a ``friend-foe" dichotomous relation. While ferromagnetism intuitively becomes a foe to spin-singlet superconductivity due to the latter's opposite spin alignments of the two electrons in the superconducting Cooper pairs, proximity to an antiferromagnetism phase can give rise to unconventional spin-singlet superconductivity \cite{Scalapino95,Hirschfeld11,sfbook}. Still, spin-singlet superconductivity can sometimes coexist with ferromagnetism by forming Cooper pairs with a finite center-of-mass momentum, resulting in finite-momentum superconductivity \cite{Fulde64,Larkin64}. 
Hybrid superconducting structures offer an additional way to combine magnetism and superconductivity, by explicitly combining materials with either magnetic and superconducting properties.
Such a setup gives the possibility of also realizing Majorana fermions through achieving topological superconductivity \cite{Alicea12,Beenakker13,Aguado17,Lutchyn18,Flensberg21}, as well as produces interesting prospects for energy efficient spintronic technologies \cite{Eschrig08,Linder15}.

Recently, a new form of magnetism, called altermagnetism, has been proposed theoretically \cite{Smejkal20,Yuan20,Mazin21,Smejkal22a,Smejkal22,Hayami19,Hayami20} and observed experimentally \cite{Feng22,Gonzalez23,Bai23,Olena24,Zhu24,Krempasky24} in a large number of materials \cite{bai24}, beyond the previously well-established ferro- and antiferromagnetic possibilities. In contrast to both ferro- and antiferromagnets, altermagnets break the Kramer's spin-degeneracy but with a momentum dependent spin splitting, still keeping zero net magnetization. Another essential feature that makes altermagnets different is that the system breaks into two spin-sublattices, which cannot be mapped onto each other by translation or inversion \cite{Mazin24}. Since the symmetry transformation that connects the two spin-sublattices does not leave the electronic spectra invariant, the electron bands end up with the characteristic momentum dependent spin splitting \cite{Mazin24}.

The most common mechanism of generating altermagnetism is through the presence of an electric crystal field splitting and spin-exchange splitting \cite{Smejkal22,Smejkal22a}. Such a mechanism does not require any Fermi surface instability into an ordered state and can thus be straightforwardly captured within first-principles density functional theory (DFT) calculations \cite{Smejkal22,Smejkal22a}. 
Alternate mechanisms for generating altermagnetism through electron-electron interactions resulting in a Fermi surface instability has also recently been proposed theoretically \cite{Maier23,Leeb24,Brekke23,Roig24,Yu24}. Such interaction-induced altermagnetism can be distinguished into two broad classes, one where electron-electron repulsion leads to an onsite magnetization, which along with a rotational symmetry breaking band structure gives altermagnetism \cite{Maier23,Leeb24,Brekke23,Roig24,Yu24}, while another mechanism dates longer back originally introduced under the name of a spin-Pomeranchuk instability, where magnetization lives on bonds \cite{Wu07}. However experimental indications of such a spin-Pomeranchuk mechanism is still lacking \cite{Smejkal22}. 

Due to the unique spin structure of altermagnets, its interplay with superconductivity was already early on anticipated to be intriguing \cite{Mazin22} and has since then generated significant attention. For example, when altermagnetism and superconductivity coexist intrinsically in the same material, it has theoretically been predicted that finite-momentum superconductivity can appear even in the absence of any net magnetization \cite{Rodrigo14, Chakraborty24, Shuntaro23, bose24, sim24, hong24}. Experimentally, hints of such coexistence also exists in the altermagnetic candidate material RuO$_{2}$ under strain \cite{Uchida20,Smejkal23} and monolayer FeSe \cite{mazin23}. Moreover, the parent cuprate material La$_{2}$CuO$_{4}$, which upon doping displays high-temperature superconductivity, is also proposed to host altermagnetism \cite{Smejkal22a}.
In a different direction, theoretical investigations considering superconductor-altermagnet hybrid junctions have also uncovered great prospects for exotic phenomena, such as intriguing Josephson effects \cite{Zhang24,Ouassou23,Papaj23,sun24,Lu24}, phase-shifted Andreev levels \cite{Beenakker23}, prospects for stray-field-free memory devices \cite{Hans24}, controllable superconducting diode effect \cite{Banerjee24}, dissipationless spin-splitting and filtering effects \cite{giil24}, magnetoelectric effect \cite{Zyuzin24}, realizing topological superconductivity \cite{Li24,Ghorashi23,Zhu23}, and spin-polarized specular Andreev reflections \cite{nagae24}. However, despite this large number of theoretical works on the interplay of altermagnetism and superconductivity, most seemingly miss to consider the crucial aspect that spin and sublattice symmetries are intrinsically coupled in altermagnets.  

In this work, we show that the strong intrinsic coupling of spin and sublattice in altermagnets has a decisive effect on the nature of superconducting pairing symmetries in altermagnets, irrespective of the origin of superconductivity. By investigating the prototype four-band model of altermagnetism, proposed in Ref.~\cite{Smejkal22a}, we first identify that the two low-energy bands relevant for superconductivity have an essentially complete spin-sublattice locking. We then uncover that this spin-sublattice locking imposes severe constraints on the superconducting symmetries in altermagnets. In particular, we find that the two most commonly studied superconducting pairing symmetries, spin-singlet onsite, or uniform, $s$-wave, and spin-triplet nearest-neighbor $p$-wave pairing are not allowed by symmetries. In fact, for a $d_{x^2-y^2}$-wave altermagnet, the most commonly studied altermagnet, we find that the only possible spin-singlet pairing must have $d_{x^2-y^2}$-wave or extended $s$-wave (i.e.~beyond onsite) symmetry, unless considering higher harmonics or angular momentum, which are both much less likely to appear. However, also allowing for unequal-time Cooper pairs, we show that further possibilities emerge with odd-frequency pairing, including onsite $s$-wave equal-spin-triplet odd-frequency pairing. We additionally explore a model for interaction-induced altermagnetism and find that its spin Fermi surfaces are also fully sublattice polarized and thus generates the same constraints for the superconducting pairing as in the prototype non-interacting four-band model. Using this interaction-induced altermagnet model, we explicitly show that our analysis of the superconducting pairing symmetries holds even when allowing for finite momentum pairing.

We organize the rest of this paper in the following way. In Sec.~\ref{sec:twoband} we discuss the prototypical non-interacting model for altermagnetism. We first show the construction of the minimal two-band model in Sec.~\ref{sec:alttwoband}. We then find the possible superconducting pairing symmetries in Sec.~\ref{sec:sctwoband}. Next, in Sec.~\ref{sec:intmodel}, we explore a different, interaction-induced altermagnet model where we first demonstrate altermagnetism within a mean-field analysis of the electronic interaction in Sec.~\ref{sec:norintmodel} and then calculate different superconducting order parameters self-consistently allowing also for finite momentum pairing in Sec.~\ref{sec:scintmodel}. Finally, we summarize our results in Sec.~\ref{sec:concl}.

\section{Band-structure altermagnetism}\label{sec:twoband}

Altermagnetism can appear in materials in two ways. First, the most studied model for altermagnetism was originally proposed in Ref.~\cite{Smejkal22a}, where a combination of electric crystal field splitting and large spin-exchange splitting generates a momentum dependent spin-splitting of the electronic band structure. Here, altermagnetism is a band structure effect and not a Fermi surface instability and can thus be captured within first-principles DFT calculations \cite{Smejkal22,Smejkal22a}. A second alternative is altermagnetism instead originating from electronic interactions through a Fermi surface instability, providing a very different but equally valid mechanism \cite{Maier23,Leeb24,Brekke23,Roig24,Yu24,Wu07}. In this section, we consider the first, non-interacting model for altermagnetism, while, later in Sec.~\ref{sec:intmodel}, we consider an interaction-induced model.

\subsection{Construction of minimal two-band model}\label{sec:alttwoband}

The original non-interacting altermagnets can be well-captured in a four-band model on the square lattice with $d_{x^2-y^2}$-wave altermagnetism  \cite{Smejkal22a},
\begin{equation}
H_{4b}=\sum_{k} \Psi^{\dagger} \hat{H}_{4b} \Psi,
\label{eq:4bandmodel}
\end{equation}
with 
\begin{eqnarray}
\hat{H}_{4b}=\left(-2t(\cos k_x+\cos k_y)-\mu\right)\sigma_{0}\tau_{0}  \nonumber \\
-\frac{t_{\rm am}}{2}(\cos k_x-\cos k_y)\sigma_{0}\tau_{z}-J\sigma_{z}\tau_z, 
\label{eq:4bandmodelm}
\end{eqnarray}
using the basis $\Psi^{\dagger}=\left(c_{kA\uparrow}^{\dagger},c_{kB\uparrow}^{\dagger},c_{kA \downarrow}^{\dagger},c_{kB\downarrow}^{\dagger}\right)$, where $c_{ k A/B \sigma}^{\dagger}$ ($c_{k A/B \sigma}$) are the creation (annihilation) operators of an electron with spin $\sigma$ and momentum $k$ on the sublattice $A/B$. Further, $t$ is the nearest-neighbor hopping amplitude, $t_{\rm am}$ denotes the electric crystal field splitting, $J$ is the momentum independent spin-exchange interaction, while $\mu$ is the chemical potential tuned to fix the average density of electrons. Here, $\sigma$ and $\tau$ are Pauli matrices in spin and sublattice space, respectively. It is important to highlight a few features of the Hamiltonian in Eq.~\eqref{eq:4bandmodelm}. First, the exchange splitting is opposite for different sublattices due to the $\tau_z$ term, which is a characteristic feature of altermagnets. Secondly, in this four-band model, $t_{\rm am}$ term is not spin-dependent, as expected for electric crystal field splitting. 

\begin{figure}[ht]
\includegraphics[width=1.0\linewidth]{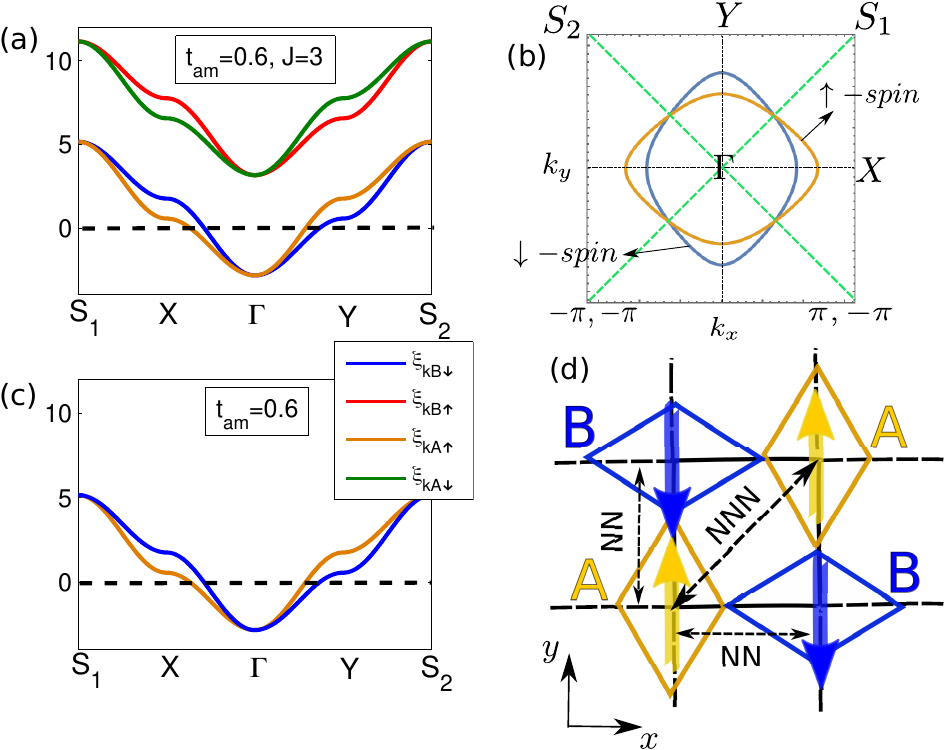} 
\caption{Energy bands in the four-band model in Eq.~\eqref{eq:4bandmodel} (a) and extracted low-energy two-band model in Eq.~\eqref{eq:2bandmodel} (c). (b) Fermi surface of both four-band and two-band models with blue denoting $\downarrow$-spin band and yellow denoting $\uparrow$-spin band with indicated high-symmetry points, $\Gamma,X,Y,S_1,S_2$, of the first BZ. Green dashed lines indicate the nodes of the altermagnet. (d) Real-space arrangement of the low-energy bands drawn on the underlying square lattice (black lines). Yellow and blue colors denote $\uparrow$-spin and $\downarrow$-spin band projections in real-space, respectively. Distortions of the diamond lattice sites A and B signify that they have different local environment and hence different crystal field splittings, i.e.~$t_{\rm am}\ne 0$. Nearest-neighbor (NN) and next-nearest-neighbor (NNN) sites are indicated by dashed double-arrows.
}
\label{fig:4bandto2band} 
\end{figure}

To understand the origin of altermagnetism, we plot in Fig.~\ref{fig:4bandto2band}(a) the four energy bands of Eq.~\eqref{eq:4bandmodel}, which we label $\xi_{kA\uparrow}$, $\xi_{kA\downarrow}$, $\xi_{kB\uparrow}$, and $\xi_{kB\downarrow}$, for $t=1$ (i.e.~all energies are in units of $t$) and a fixed average electron density $\rho=0.6$. Since the Hamiltonian in Eq.~\eqref{eq:4bandmodel} is diagonal, the four bands can be labeled by the same sublattice and spin indices as in the chosen $\Psi$ basis. This  basis and labeling of the four bands using sublattice and spin indices is motivated by the findings of Ref.~\cite{Smejkal22a}. For $t_{\rm am}=0$ and $J=0$, all the four bands are degenerate. With finite $t_{\rm am}=0.6$, but still $J=0$, the A and B bands split, indicating sublattice symmetry breaking, but there is still spin-degeneracy. Then, also a finite $J=3$ results in spin symmetry breaking. We here consider $J$ to be large, just as in Ref.~\cite{Smejkal22a}.  Hence a momentum dependent spin-splitting appears for both finite $t_{\rm am}$ and $J$. 
As seen in Fig.~\ref{fig:4bandto2band}(a), the spin-splitting occurs in opposite direction for the two different sublattices. Consequently, the two lowest energy bands become $\xi_{kA\uparrow}$ and $\xi_{kB\downarrow}$ with spin and sublattice degree of freedoms intrinsically coupled.
The corresponding Fermi surfaces in the first Brillouin zone (BZ) are shown in \ref{fig:4bandto2band}(b), which also illustrates the nodal lines of the altermagnet (green dashed lines), where the band spin-splitting is zero. The other two bands, $\xi_{kA\downarrow}$ and $\xi_{kB\uparrow}$, are split off from these two low-energy states by an energy scale of $J=3$, which is known to be large \cite{Smejkal22a}. 
 
 For the purpose of considering superconducting pairing it is particularly important to here note that different spin bands are associated to different sublattices. The main reason behind this is that, as shown in Ref.~\cite{Smejkal22a}, for altermagnetic materials the yellow low-energy $\uparrow$-spin (and green high-energy $\downarrow$-spin) band in Fig.~\ref{fig:4bandto2band}, when projected in real-space, have maximum weight on sublattice A, while the blue low-energy $\downarrow$-spin (and red high-energy $\uparrow$-spin) band, when projected in real-space, have maximum weight on sublattice B. We schematically illustrate this in the real-space schematic in Fig.~\ref{fig:4bandto2band}(d). The differently distorted diamonds on the square lattice (black lines) indicate the different local crystal field environments for sublattices A and B, set by the $t_{\rm am}$ term in Eq.~\eqref{eq:4bandmodelm}.

With only two bands close to the Fermi level, we can, for any low-energy phenomena and particularly superconductivity, to a very good approximation consider an effective two-band model consisting of only the energy bands $\xi_{kA\uparrow}$ and $\xi_{kB\downarrow}$. 
 The corresponding two-band Hamiltonian is given by
\begin{equation}
H_{2b}=\sum_{k} \Psi_{2b}^{\dagger} \hat{H}_{2b} \Psi_{2b},
\label{eq:2bandmodel}
\end{equation}
with 
\begin{equation}
\hat{H}_{2b}=\left(-2t(\cos k_x+\cos k_y)-\mu\right)s_{0}-\frac{t_{\rm am}}{2}(\cos k_x-\cos k_y)s_{z},
\label{eq:2bandmodelm}
\end{equation}
and now using the basis $\Psi_{2b}^{\dagger}=\left(c_{kA\uparrow}^{\dagger},c_{kB\downarrow}^{\dagger}\right)$. Note that here $s$ now is a $2\times 2$ Pauli matrix in the joint sublattice and spin basis given by $\Psi_{2b}$. In this two-band model $t_{\rm am}$ now plays the role of momentum dependent spin-splitting with nodes along $|k_x|=|k_y|$. The energy bands of Eq.~\eqref{eq:2bandmodel} are shown in Fig.~\ref{fig:4bandto2band}(c). These bands are exactly same as the two low-energy bands of Fig.~\ref{fig:4bandto2band}(a) by construction, and hence give the same Fermi surface as in Fig.~\ref{fig:4bandto2band}(b).

\begin{table*}[t]
\begin{center}
\begin{tabular}{ |c|c|c|c|c| } 
\hline
Spin symmetry ($S$) & Momentum symmetry ($P$) & Form factor & Real-space structure & Occurrence \\
\hline
\hline
\multirow{4}{10em}{Spin-singlet ($S=-1)$} & $s$-wave ($P=1$) & 1 & Onsite & \large{\textcolor{red}{$\pmb\times$}} \\ 
& Extended $s$-wave ($P=1$) & $\cos{k_x}+\cos{k_y}$ & NN & \large{\textcolor{green}{$\checkmark$}} \\ 
&  $d_{x^2-y^2}$-wave ($P=1$) & $\cos{k_x}-\cos{k_y}$ & NN & \large{\textcolor{green}{$\checkmark$}} \\ 
& $d_{xy}$-wave ($P=1$) & $\sin{k_x}\sin{k_y}$ & NNN & \large{\textcolor{red}{$\pmb\times$}} \\ 
\hline
\multirow{1}{12em}{Equal-spin-triplet ($S =1$)} & $p$-wave ($P=-1$) & $\sin{k_x}$, $\sin{k_y}$ & NN & \large{\textcolor{red}{$\pmb\times$}} \\ 
\hline
\multirow{1}{12em}{Mixed-spin-triplet ($S =1$)} & $p$-wave ($P=-1$) & $\sin{k_x}$, $\sin{k_y}$ & NN & \large{\textcolor{green}{$\checkmark$}} \\ 
\hline
\end{tabular}
\end{center}
\caption{Possible occurrence of superconducting pairing symmetries in a $d_{x^2-y^2}$-wave altermagnet on a square lattice, Eq.~\eqref{eq:4bandmodel}, indicated by their spin symmetry, momentum structure or parity, corresponding form factor, and simplest real-space structure, considering up to NNN pairing in real space. 
Only equal-time (or equivalently even-frequency) BCS-like pairings are shown.}
\label{tab:table1}
\end{table*}

\subsection{Superconducting pairing}\label{sec:sctwoband}

Since superconductivity is a low-energy phenomenon, it is sufficient to consider only the two bands closest to the Fermi level, as given by the Hamiltonian Eq.~\eqref{eq:2bandmodel}. In most of the literature on superconductivity in altermagnets, the sublattice index in Eq.~\eqref{eq:2bandmodel} is dropped. However, we should be careful with this sublattice index when discussing superconductivity, since both spatial and spin symmetries are crucial for determining the possible superconducting pairing symmetries \cite{Sigrist91}. Superconductivity occurs due to the formation of Cooper pairs made of two electron-like quasiparticles. Because of the fermionic nature of the constituents, there are restrictions on the nature of the Cooper pairs. In particular, under a joint operation of spin permutation ($S$), momentum exchange or parity ($P$), and relative time permutation ($T$) of the individual electrons, superconducting pairing needs to satisfy the condition $SPT=-1$, for a single orbital system \cite{Triola16,Linder19,Triola20} \footnote{Alternatively, we may treat the sublattice index as an orbital degree of freedom \cite{Black-Schaffer13} within an extended unit cell consisting of both an A and B site. This results in instead using the condition $SPOT =-1$, where $O$ stands for orbital parity, while the spatial parity becomes $P=1$ within the unit cell. This results in the same conclusions as in Tables \ref{tab:table1}-\ref{tab:table2} but the resulting spatial symmetries (form factors) are less straightforward to extract.}.

Let us begin our discussion on the pairing symmetries possible in a band-structure generated altermagnet, Eq.~\eqref{eq:2bandmodel} by first considering only equal-time pairing, as is the standard (BCS)  form of superconductivity. Equal-time pairing automatically means $T=1$, since there is no consequences of exchanging the time coordinate of the individual electrons in the Cooper pair. In this scenario, only spin and spatial symmetries decide the nature of the pairing. Then, in order to satisfy $SPT=-1$, spin-singlet pairing ($S=-1$) has to be even-parity in $k$-space ($P=1$), while spin-triplet pairing ($S=1$) has to be odd-parity ($P=-1$), as is standard for superconductivity.

\subsubsection{Spin-singlet pairing}
Focusing first on spin-singlet pairing, the simplest, and most common, is a uniform or isotropic $s$-wave symmetry (also known as conventional) pairing. This leads to pairing of opposite spins and on the same sublattice, since the uniformness of the $s$-wave pairing in $k$-space leads to onsite pairing in real space. However, in Eq.~\eqref{eq:2bandmodel} and explicitly shown in Fig.~\ref{fig:4bandto2band}, opposite spins of the same sublattice are not present in the two-band minimal model. Consequently, onsite $s$-wave pairing is not allowed in altermagnets. 
If we next expand and consider spin-singlet pairing on nearest-neighbor (NN) sites in real space, the possible spin-singlet candidates are $d_{x^2-y^2}$-wave and extended (i.e.~beyond onsite) $s$-wave, given by the form factors in Table \ref{tab:table1}. As NN sites belong to different sublattices and thus have opposite spins, both of these pairing symmetries are allowed in an altermagnet. 
Further considering next-nearest-neighbor (NNN) spin-singlet pairing, we recover also the $d_{xy}$-wave symmetry. However, since NNN pairing occurs between diagonal sites on the square lattice, which, as shown in Fig.~\ref{fig:4bandto2band} carry equal spins in an altermagnet, spin-singlet $d_{xy}$-wave pairing is not allowed. 
Consequently, the only possible spin-singlet pairing symmetries up to NNN in real space are the $d_{x^2-y^2}$-wave and extended $s$-wave symmetries. We summarize these results in Table \ref{tab:table1}. Considering longer-range pairing in real space results in either (specific) higher harmonics of the same symmetries or higher order symmetries (e.g.~$g,i$-wave symmetries), all of which contain more nodes in the BZ and are thus often energetically less favorable as the superconducting condensation energy decreases when more low-energy quasiparticles are present. We thus conclude that for spin-singlet superconducting pairing, we should generally only consider $d_{x^2-y^2}$-wave and extended $s$-wave symmetries in the band-structure model, Eq.~\eqref{eq:4bandmodel}, of altermagnets. 
This explicitly excludes the most common form of superconductivity, the spin-singlet uniform $s$-wave state, commonly assumed to exist for all phonon-driven superconductors. At the same time, it allows for the $d_{x^2-y^2}$-wave state present in the high-temperature cuprate superconductors \cite{Scalapino95}. Interestingly, at least the parent cuprate material La$_{2}$CuO$_{4}$ has also been proposed to host altermagnetism \cite{Smejkal22a}.
We also note that the $d_{x^2-y^2}$-wave has the same nodal structure as the $d_{x^2-y^2}$-wave nodes of altermagnetism in Eq.~\eqref{eq:4bandmodel}. 

\subsubsection{Spin-triplet pairing}
Next, considering spin-triplet pairing, it can be of two forms, equal-spin-triplet and mixed-spin-triplet pairing depending on the spin projection along the spin axis. Equal-spin-triplet pairing would necessarily mean pairing A-sublattice with A-sublattice in the low-energy two band model, Eq.~\eqref{eq:2bandmodel}. This would be possible if the pairing occurs onsite, but this would  mean $P=1$ and $S=1$ pairing which does not satisfy $SPT=-1$ and is thus not allowed. The only possibility to achieve equal-spin-triplet pairing is to construct pairing between different A sublattice sites, because the standard NN $p$-wave pairing is blocked due to the spin structure. The shortest range possibility is pairing on second nearest neighbors, or 2x- or 2y-bonds in Fig.~\ref{fig:4bandto2band}(d). However, this is longer-range pairing compared to both NN and NNN pairing and will therefore contain more nodes in $k$-space, leading often to less condensation energy. Or alternatively, viewed in real space, longer-range pairing requires longer range interactions, which tend to be decaying with distance, such that they become subdominant. As a consequence, we deem equal-spin pairing to be unlikely in an altermagnet.
The only possible spin-triplet pairing is therefore a mixed-spin-triplet state. This state can occur on NN sites, and then have a $p$-wave symmetry. We summarize all possibilities for spin-triplet pairing up to NNN  in Table.~\ref{tab:table1}.

\subsubsection{Odd-frequency pairing}
We finally discuss the scenario where also unequal-time pairing is considered, i.e.~the two electrons can also pair at unequal times, $t_1$ and $t_2$. If the pairing is even in the relative time coordinate, $t_1-t_2$, then $T=1$ and the possible pairing symmetries are exactly same as discussed in the previous two subsections and summarized in Table \ref{tab:table1}. However, if we also allow for the pairing to be odd in the relative time coordinate, or equivalently odd in frequency pairing, additional distinct possibilities emerge due to $T=-1$. Such odd-frequency pairing \cite{Berezinskii74, Kirkpatrick91, Balatsky92, Schrieffer94, Bergeret05, Yokoyama11, Black-Schaffer13, Alidoust14, Tanaka12, Linder19, Triola20, Chakraborty21,Chakraborty22a,Chakraborty22b} has been studied extensively in superconductor-ferromagnet hybrid junctions \cite{Bergeret01,Bergeret01prb,Buzdin05,Eschrig08,DiBernardo15,Bernardo15,Jacobsen16,Ouassou17,Perrin20}, also including spintronics applications \cite{Eschrig08,Linder15}. For superconductor-altermagnet hybrid junctions, the spintronics applications can be even more technologically beneficial as the zero net magnetization in altermagnets result in no stray fields \cite{Smejkal22}. Hence, we also explore the possible superconducting pairing symmetries for odd-frequency pairing. 

Due to the oddness in relative time symmetry $T=-1$, $SPT=-1$ can now be satisfied for spin-singlet pairing ($S=-1$) with odd-parity ($P=-1$) or spin-triplet pairing ($S=1$) with even-parity ($P=1$). Consequently, the even-parity pairing symmetries not allowed with equal-time pairing, onsite $s$-wave and $d_{xy}$-wave pairing, are now allowed with an equal-spin-triplet configuration. The other even-parity pairings previously considered, $d_{x^2-y^2}$-wave and extended $s$-wave, will now instead appear in a mixed-spin-triplet configuration, due to them connecting different sublattices. 
Finally, spin-singlet, odd-parity is also allowed as a NN $p$-wave pairing.
We summarize these possible odd-frequency pairing symmetries in Table.~\ref{tab:table2}.

\begin{table}[t]
\begin{center}
\begin{tabular}{ |c|c| } 
\hline
\multicolumn{2}{|c|}{\textbf{Possible odd-frequency pairings ($T=-1$)}} \\
\hline
Spin symmetry ($S$) & Momentum symmetry ($P$) \\
\hline
\hline
\multirow{1}{10em}{Spin-singlet ($S=-1$)} & $p$-wave ($P=-1$) \\ 
\hline
\multirow{2}{12em}{Equal-spin-triplet ($S=1$)} & $s$-wave ($P=1$)  \\ 
& $d_{xy}$-wave ($P=1$)  \\ 
\hline
\multirow{2}{12em}{Mixed-spin-triplet ($S=1$)} & $d_{x^2-y^2}$-wave ($P=1$)  \\ 
& Extended $s$-wave ($P=1$)  \\ 
\hline
\end{tabular}
\end{center}
\caption{Possible occurrence of odd-frequency superconducting pairing symmetries in a $d_{x^2-y^2}$-wave altermagnet on a square lattice, Eq.~\eqref{eq:4bandmodel}, indicated by their spin symmetry and momentum structure or parity, considering up to NNN pairing in real space.}
\label{tab:table2}
\end{table}

To summarize, in this section we show that the coupled nature of spin and sublattice, essential in generating the momentum dependent spin-splitting in non-interacting altermagnets,  puts severe constraints on possible superconducting pairing, as summarized in Table \ref{tab:table1} and extended in Table \ref{tab:table2} for odd-frequency superconducting pairing. The results in this section stem from general symmetry arguments of altermagnets and superconducting pairing and thus do not assume any specific form for the superconducting pairing mechanism. Hence, the analysis is applicable to any type of intrinsic superconductivity as well as for proximity-induced superconducting pairing in hybrid structures. 

\section{Interaction-induced altermagnetism}\label{sec:intmodel}

In the last section, we consider a model for altermagnetism where the band structure alone gives rise to the characteristic momentum dependent spin-splitting. There also exist other models where altermagnetism is alternatively generated by electronic interactions \cite{Maier23,Leeb24,Brekke23,Roig24,Yu24,Wu07}. In this section, we study one such model where electronic interactions combined with band structure effects can generate altermagnetism \cite{Maier23}. However, the discussion of this section is also applicable to other interaction-induced models of altermagnetism.

\subsection{Normal state}\label{sec:norintmodel}

Following Ref.~[\onlinecite{Maier23}], we again consider the square lattice with two sublattices, but now with the non-interacting, or kinetic energy given by
\begin{eqnarray}
H_{0} = &\sum_{k,\sigma} \xi_{AA}(k) c_{ k A \sigma}^{\dagger} c_{ k A \sigma} 
+ \xi_{BB}(k) c_{k B \sigma}^{\dagger} c_{k B \sigma} \nonumber \\
&+ \,\,  \epsilon_{AB}(k) ( c_{k A \sigma}^{\dagger} c_{k B \sigma} + \textrm{H.c.} ),
\label{eq:nonint}
\end{eqnarray}
where again $c_{k A/B \sigma}^{\dagger}$ ($c_{k A/B \sigma}$) are the creation (annihilation) operators of an electron with spin $\sigma$ and momentum $k$ on the sublattice $A/B$, and with
\begin{align}
\xi_{AA}(k) & =-2t_1\cos(2k_x)-2t_2\cos(2k_y)-\mu, \nonumber \\
\xi_{BB}(k) & =-2t_2\cos(2k_x)-2t_1\cos(2k_y)-\mu, \nonumber \\
\epsilon_{AB}(k) &=-2t(\cos(k_x)+\cos(k_y)).
\label{eq:dis}
\end{align}
Here $t$ is the NN hopping amplitude, $t_{1/2}$ are second neighbor hopping amplitudes, and $\mu$ is the chemical potential tuned to fix the average density of electrons. We then add interactions in the form of an onsite repulsive Hubbard interaction,
\begin{equation}
H_{U}=\sum_{i} Un_{i\uparrow}n_{i\downarrow},
\label{eq:HUbb}
\end{equation}
where $n_{i \sigma}$ is the number operator and $U$ is the onsite Coulomb repulsion. A mean-field decoupling of the Hubbard interaction in the Hartree channel in order to achieve magnetism gives
\begin{equation}
H_{U,mf}=\sum_{i \sigma}\frac{1}{2}U\left( \rho n_{i \sigma} - m_i \sigma n_{i \sigma} \right) + \text{constant},
\label{eq:HUbbmf}
\end{equation}
where $\rho$ is the average electron density per sublattice and $m_{i}=\langle n_{i\uparrow}- n_{i\downarrow}\rangle$ is the magnetization. We here assume a homogeneous solution of the form $m_{i}=m_{A}=-m_{B}\equiv m$, which generates a sublattice dependence of the magnetization, and, if non-zero, gives either antiferromagnetism or altermagnetism.

The total Hamiltonian becomes $H_{int}=H_0+H_{U,mf}$ and can be written in a matrix form using the same basis as in Section \ref{sec:twoband}, $\Psi^{\dagger}=\left(c_{k A \uparrow}^{\dagger},c_{k B \uparrow}^{\dagger},c_{k A \downarrow}^{\dagger},c_{k B \downarrow}^{\dagger}\right)$, as
\begin{equation}
H_{int}=\sum_{k} \Psi^{\dagger} \hat{H}_{int} \Psi,
\label{eq:intmodelnormal}
\end{equation}
with 
\begin{eqnarray}
\hat{H}_{int}&=&\left(\xi_{AA}(k)+\frac{1}{2}\xi_{BB}(k)\right)\sigma_{0}\tau_{0} \nonumber \\
&+&\left(\xi_{AA}(k)-\frac{1}{2}\xi_{BB}(k)\right)\sigma_{0}\tau_{z} \nonumber \\
&+&\epsilon_{AB}(k)\sigma_{0}\tau_{x}-M\sigma_{z}\tau_{z},
\label{eq:intmodelnormalm}
\end{eqnarray}
where we define $M=\frac{1}{2}Um$, while the spin-independent Hartree shift term $\frac{1}{2}U\rho$ of Eq.~\eqref{eq:HUbbmf} is absorbed into an effective $\mu$. As shown in Ref.~[\onlinecite{Maier23}], altermagnetism appears in Eq.~\ref{eq:intmodelnormal} for finite $M$ and when there is a sublattice asymmetry with $\xi_{AA}(k) \ne \xi_{BB}(k)$. From Eq.~\eqref{eq:dis}, we see that the condition $\xi_{AA}(k) \ne \xi_{BB}(k)$ is equivalent to $t_1 \ne t_2$. 

Beyond the same basis, Eq.~\eqref{eq:intmodelnormalm} also has a very similar matrix form to Eq.~\eqref{eq:4bandmodelm} for band-structure generated altermagnetism. The magnetization $M$ in Eq.~\eqref{eq:intmodelnormalm} plays the role of the spin-exchange splitting $J$ in Eq.~\eqref{eq:4bandmodelm}, although $M$ is now an order parameter obtained self-consistently from a Hubbard interaction $U$. Here, $t_1 - t_2$ plays the role of the crystal field splitting $t_{\rm am}$ in Eq.~\eqref{eq:4bandmodelm}. However, in contrast to Eq.~\eqref{eq:4bandmodelm}, Eq.~\eqref{eq:intmodelnormalm} has an additional hybridization between the two sublattices through the term $\epsilon_{AB}(k)\sigma_{0}\tau_{x}$. Because of this hybridization, Eq.~\eqref{eq:intmodelnormalm} is no longer diagonal and hence the four diagonal bands cannot automatically be labeled by the same indices as in the basis $\Psi$. However, as we will see later, the spin bands will still be strongly coupled to sublattice in the altermagnetic state, just as in Eq.~\eqref{eq:4bandmodelm}.

\begin{figure}[t]
\includegraphics[width=1.0\linewidth]{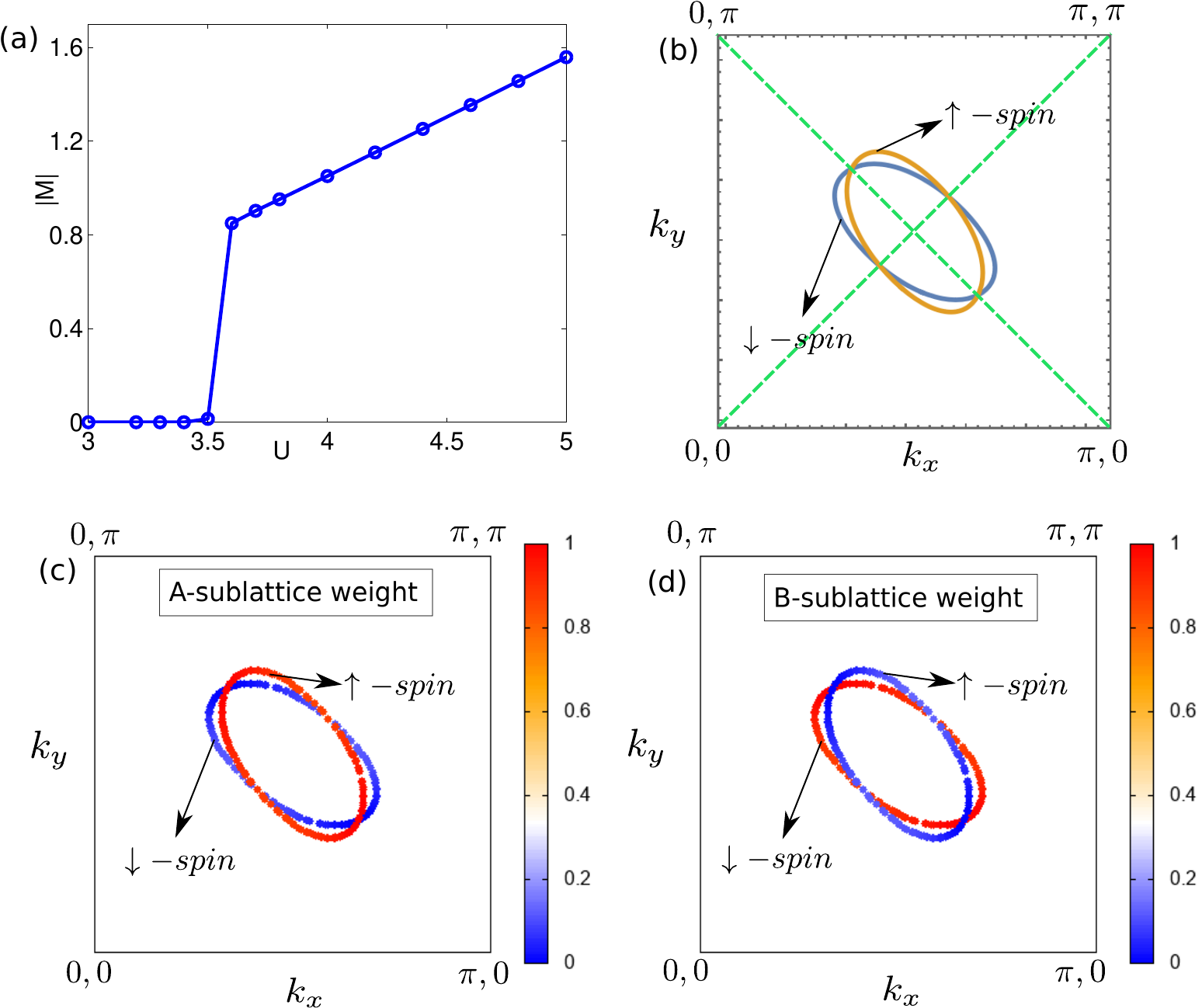} 
\caption{(a) Magnetization per sublattice $|M|$ as a function of $U$ for the interaction-driven altermagnet model in Eq.~\eqref{eq:intmodelnormal} with extracted Fermi surface for $U=4.6$ (b). Green dashed lines in (b) indicate the nodes of the altermagnet. (c,d) Normalized sublattice weight calculated for the low-energy bands crossing the Fermi surface. In (b-d) only the positive quadrant of the first BZ is shown for a better visibility. Other quadrants are obtained by four-fold rotational symmetry.
 }
\label{fig:interactionres} 
\end{figure}

To proceed with finding interaction-driven altermagnetism, we solve the Hamiltonian Eq.~\eqref{eq:intmodelnormal} self-consistently for the order parameter $M$. In the following we report results for $t_1-t_2=0.2$, using again $t=1$ as the energy unit, and for a fixed average density per sublattice $\rho=0.9$ and we use a system size of $1000\times 1000$ $k$-points in momentum space. Larger system sizes do not change the results.

In Fig.~\ref{fig:interactionres}(a) we show the obtained value of $|M|$ with varying $U$. For $U\le 3.5$, there is no magnetization, $|M|=0$. At a critical $U=3.6$, $|M|$ acquires a finite large value and then $|M|$ increases linearly with increasing $U$. The corresponding Fermi surfaces in the ordered phase are shown in \ref{fig:interactionres}(b) in the positive quadrant of the first BZ. Due to four-fold rotational symmetry each quadrant hold a similar Fermi surface.
Due to the finite $|M|$ and asymmetry $t_1-t_2$, the Fermi surfaces of $\uparrow$-spin (yellow) and $\downarrow$-spin (blue) are split, except along the nodal lines $|k_x|=|k_y|$ and $|k_x\pm k_y|=\pi$ (green dashed). These results establish that a finite interaction $U$ generates altermagnetism with nodes along both $|k_x|=|k_y|$ and $|k_x\pm k_y|=\pi$, the latter generating additional nodal lines compared to the band-structure altermagnetism in Sec.~\ref{sec:twoband}. For $t_1-t_2=0$, a finite $|M|$ instead generates antiferromagnetism, with a non-split Fermi surface (not shown). We further comment on the value of the model parameters. For a fixed $\rho$, a larger (smaller) $t_1-t_2$ simply results in a larger (smaller) critical $U$. Moreover, our choice of $\rho=0.9$ is motivated by two reasons. First, at half filling per sublattice, $\rho=1.0$, the altermagnetic state is metallic only for a very narrow window of $U$ \cite{Maier23}. To get superconductivity a metallic altermagnetic state is required and we therefore choose $\rho=0.9\ne 1.0$ in order to achieve a doped metallic altermagnet for a larger window of $U$. Secondly, filling levels $\rho \ll 1.0$ and $\rho \gg 1.0$ are also detrimental to interaction-induced altermagnetism since such densities need very large $U$ to get finite $M$.

As mentioned above, in contrast to in Sec.~\ref{sec:twoband}, Eq.~\eqref{eq:intmodelnormalm} is no longer a diagonal matrix. Hence, in order to identify the superconducting symmetries it is important to find out the sublattice weights of the low-energy bands in this model as they are not necessarily inherited from the original basis $\Psi$. In Figs.~\ref{fig:interactionres}(c-d) we plot the normalized projection of the eigenvectors of the two low-energy diagonal bands on the A-sublattice (c) and the B-sublattice (d). We only show the sublattice weights for $k$-points on the Fermi surface since only the Fermi surface is essential to superconductivity. As seen, the $\uparrow$-spin Fermi surface is purely A-sublattice polarized, while the $\downarrow$-spin Fermi surface is purely B-sublattice polarized. This establishes that even an interaction-induced model for altermagnetism has the same basic properties or spin-sublattice locking as discussed in the archetype four-band model of Sec.~\ref{sec:twoband}.

\subsection{Superconducting pairing}\label{sec:scintmodel}

Due to the same nature of spin-sublattice locking in the interaction-induced altermagnet model Eq.~\eqref{eq:intmodelnormal} as in the band-structure induced altermagnet model Eq.~\eqref{eq:4bandmodel}, all the discussion of superconducting pairing symmetries in Sec.~\ref{sec:sctwoband}, with results summarized in Tables \ref{tab:table1},\ref{tab:table2}, is also valid for the interaction-induced model, with exactly same constraints on the pairing symmetries. In this section, we demonstrate this explicitly by calculating three different examples of spin-singlet pairing, uniform $s$-wave, extended $s$-wave, and $d_{x^2-y^2}$-wave (called $d$-wave pairing from here on) symmetries, and now also by allowing for the possibility of finite-momentum superconductivity.

To be able to investigate superconductivity, we add to Eq.~\eqref{eq:intmodelnormal} a NN attractive interaction $V_{sc}$ for generating $d$-wave and extended $s$-wave pairing, and an onsite attractive interaction $U_{sc}$ for generating onsite $s$-wave pairing,
\begin{equation}
H_{sc}=\sum_{\langle ij \rangle\sigma\sigma'} -V_{sc} n_{i\sigma}n_{j\sigma'}-\sum_{i} U_{sc} n_{i\uparrow}n_{i\downarrow},
\label{eq:nnscterm}
\end{equation} 
We next mean-field decouple $H_{sc}$ in the spin-singlet Cooper channel and not in the normal channel. Primary motivation behind this is that the origin of $H_{sc}$ in Eq.~\eqref{eq:nnscterm} and $H_{U}$ in Eq.~\eqref{eq:HUbb} can be very different. For example, an onsite attraction in Eq.~\eqref{eq:nnscterm} often effectively arises due to electron-phonon coupling, whereas $H_{U}$ is purely stemming from Coulomb repulsion. Upon mean-field decoupling $H_{sc}$ in the spin-singlet Cooper channel, allowing for a finite center-of-mass momentum of the Cooper pairs $Q$, we arrive at \cite{Chakraborty24}
\begin{eqnarray}
H_{sc,mf}&=& \sum_{k} \left( \Delta^{Q}_{k} c_{ (-k+Q/2) A \downarrow} c_{ (k+Q/2) B \uparrow} + \textrm{H.c.} \right) \nonumber \\
&&+ \left( \Delta^{Q}_{s} c_{(-k+Q/2) A \downarrow} c_{(k+Q/2) A \uparrow} \right. \nonumber \\
&&\left.+\Delta^{Q}_{s} c_{(-k+Q/2) B \downarrow} c_{(k+Q/2) B \uparrow} + \textrm{H.c.} \right) \nonumber \\
&&+ \text{ constant},
\label{eq:nnsctermmf}
\end{eqnarray}
where $\Delta^{Q}_{k}$ and $\Delta^{Q}_{s}$ are NN and onsite spin-singlet superconducting order parameters, respectively, obtained from the self-consistency relations 
\begin{eqnarray}
&&\Delta^{Q}_k=\sum_{k^{\prime}}\frac{1}{2}V_{k,k^{\prime}} \langle c_{(k^{\prime}+Q/2) A \uparrow}^{\dagger} c_{(-k^{\prime}+Q/2) B \downarrow}^{\dagger} +\textrm{H.c.} \rangle, \nonumber \\
&&\Delta^{Q}_{s}=-\sum_{k^{\prime}}U_{sc} \langle c_{(k^{\prime}+Q/2) A/B \uparrow}^{\dagger} c_{(-k^{\prime}+Q/2) A/B \downarrow}^{\dagger} \rangle.
\label{eq:scsc}
\end{eqnarray}
Here the interaction $V_{k,k^{\prime}}$ is given by
\begin{equation}
V_{k,k^{\prime}}=-V_{sc}\left(\gamma(k)\gamma(k')+\eta(k)\eta(k')\right), \label{eq:int2} 
\end{equation}
with $\gamma(k)=\cos(k_x)+\cos(k_y)$ and $\eta(k)=\cos(k_x)-\cos(k_y)$ being the two form factors for NN interaction on a square lattice. 
 Incorporating the momentum dependence of $V_{k,k^{\prime}}$, we can finally write $\Delta^{Q}_k=\Delta^{Q}_d\eta(k)+\Delta^{Q}_{ext-s}\gamma(k)$, with $\Delta^{Q}_d$ being the $d$-wave superconducting order parameter and $\Delta^{Q}_{ext-s}$ being the extended $s$-wave superconducting order parameter \cite{SudboBook}. Here, the nodes of the $d$-wave state lie along the $k_x=\pm k_y$ lines which partially matches with the altermagnet nodes obtained in Fig.~\ref{fig:interactionres}(b).
 By allowing for a finite $Q$, finite-momentum pairing becomes a possibility, while $Q=0$ corresponds to (zero-momentum) BCS pairing. Note that all the three superconducting order parameters, onsite $\Delta^{Q}_{s}$, $\Delta^{Q}_d$, and $\Delta^{Q}_{ext-s}$, parametrically depend on $Q$.
 
We formulate the total Hamiltonian $H=H_{0}+H_{U,mf}+H_{sc,mf}$ using Eqs.~\eqref{eq:nonint}, \eqref{eq:HUbbmf}, and \eqref{eq:nnsctermmf} in a Nambu basis and solve self-consistently for both the superconducting order parameters $\Delta^{Q}_{s}$, $\Delta^{Q}_d$, and $\Delta^{Q}_{ext-s}$, as well as the normal state order parameter $M$ for a particular $Q$ and fixed $\rho$. Then, we obtain the true ground state by minimizing the ground state energy with respect to $Q$. This allows us to capture any putative finite-momentum pairing state. For simplicity, we use the same parameters as in Sec.~\ref{sec:norintmodel}, i.e.~$\rho=0.9$, $t_1-t_2=0.2$, and a system size of $1000\times 1000$.

\begin{figure}[t]
\includegraphics[width=1.0\linewidth]{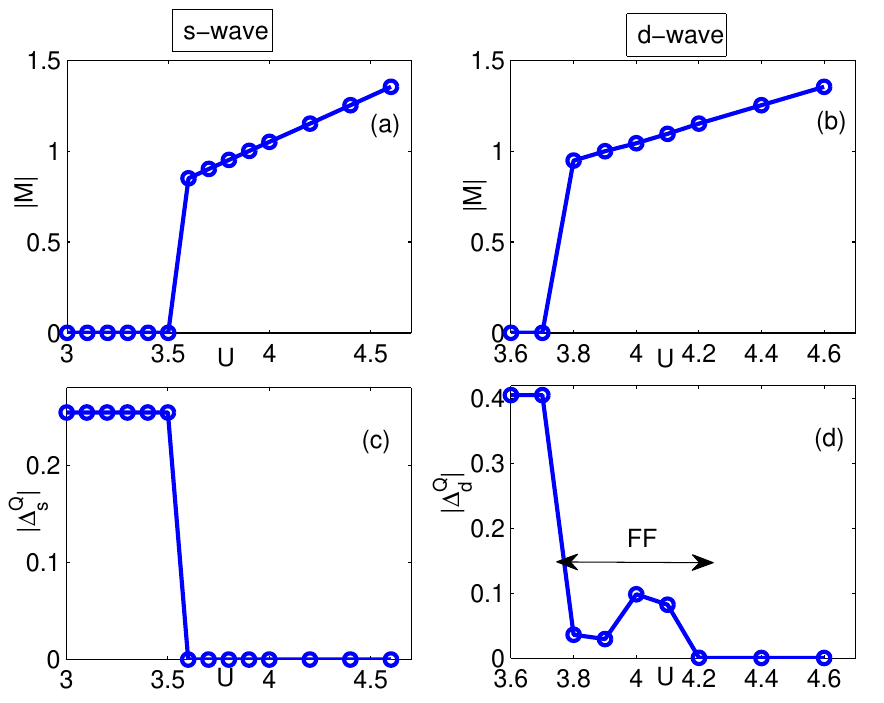} 
\caption{Magnetization per sublattice $|M|$ (a,b), superconducting order parameters $\Delta^{Q}_{s}$ (c) and $\Delta^{Q}_{d}$ (d) as a function of $U$. In (a,c), only onsite $s$-wave pairing is considered by setting $U_{sc} = 2$ and $V_{sc} =0$, whereas in (b,d) NN pairing is considred by setting $U_{sc} = 0$ and $V_{sc} =2$. 
 }
\label{fig:selfconscop} 
\end{figure}

We start by investigating at the possibility of uniform $s$-wave pairing by setting $U_{sc}=2$ and $V_{sc}=0$. In Figs.~\ref{fig:selfconscop}(a,c), we plot $|M|$ and $\Delta^{Q}_s$ as a function of varying $U$. Note that it is not necessary that the behavior of $|M|$ in Fig.~\ref{fig:selfconscop}(a) will be same as in Fig.~\ref{fig:interactionres}(a) due to the possible competition with $\Delta^{Q}_s$. However, despite this, we find the same behavior of $|M|$ with $U$ as in Fig.~\ref{fig:interactionres}(a). In terms of $\Delta^{Q}_s$, we find it to only be finite for $U\le 3.5$, when also $|M|=0$. In fact, $\Delta^{Q}_s$ sharply goes to zero as soon as $|M|$ acquires a finite value. Hence, onsite $s$-wave pairing does not survive in the presence of altermagnetism. This explicitly demonstrates that the spin-sublattice locking in the altermagnet state, seen in Fig.~\ref{fig:interactionres}, entirely prevents developing an onsite $s$-wave pairing, a result also fully consistent with the results in Table~\ref{tab:table1}. Note that this result also holds for all finite-momentum $s$-wave superconducting states.

Next, to show that NN spin-singlet pairing can occur in altermagnets, we focus on $d$-wave and extended $s$-wave pairings by setting $V_{sc}=2$ and $U_{sc}=0$. In Fig.~\ref{fig:selfconscop}(b), we first show $|M|$ as a function of $U$. We find that with the competition of NN pairing, the critical $U$ for altermagnetism to appear is shifted to $U=3.8$, i.e.~slightly greater than the critical $U$ in Fig.~\ref{fig:interactionres}(a).
For the superconducting order we find in Fig.~\ref{fig:selfconscop}(d) a large $\Delta^{Q}_d$ for $U\le 3.7$ with $Q=0$ as the ground state, i.e.~where $|M|=0$. Within the altermagnetic phase marked by a finite $|M|$, $\Delta^{Q}_d$ still survives over a range $3.7<U<4.4$, but now the ground state is at finite $Q$, meaning a finite-momentum Fulde-Ferrell (FF) state \cite{Fulde64} is formed. This result establishes that NN $d$-wave pairing is feasible in altermagnets, a result again fully in line with Table~\ref{tab:table1}. Further increasing $U$ eventually results in $\Delta^{Q}_d=0$ due to a large value of $|M|$.
We further find that the obtained values of $\Delta^{Q}_{ext-s}$ are always very small compared to $\Delta^{Q}_d$ (not shown). We note that this smallness of $\Delta^{Q}_{ext-s}$ is not related to altermagnetism but persists also when $|M|=0$. Instead, the relative strength of $\Delta^{Q}_{ext-s}$ and $\Delta^{Q}_d$ depends on the density-dependent Fermi surface, and for the chosen value of $\rho=0.9$ per sublattice in this work, $\Delta^{Q}_{ext-s}$ is small. Hence, with a different $\rho$, we have checked that a finite $\Delta^Q_{ext-s}$ can also be achieved.
We note that we here choose to not explicitly calculate results for NNN $d_{xy}$-wave or spin-triplet pairing symmetries, in order to avoid unnecessary computational complexity resulting from larger matrices and multiple order parameters. However, the similarities of our results in this section and Table~\ref{tab:table1} suggest that these pairings will also not appear if calculated self-consistently.

To summarize, in this section we show that interaction-induced altermagnetism also exhibits a spin-sublattice locking, similar to the one present in the band-structure or non-interacting model of Sec.~\ref{sec:twoband}. Hence, we find exactly the same fundamental constraints on superconducting pairing as in Sec.~\ref{sec:twoband}. We further explicitly demonstrate these constraints by self-consistently calculating the spin-singlet superconducting order parameters up to NNs along with the magnetization responsible for altermagnetism, also allowing for finite-momentum pairing. Our results demonstrate that the constraints on superconducting pairing symmetries established in Table~\ref{tab:table1} are fully generic and apply to widely different models of altermagnetism, including considering finite-momentum pairing.

\section{Conclusions}\label{sec:concl}

In conclusion, in this work we highlight the importance of the coupled spin and sublattice character of the low-energy bands in altermagnets by exploring two different models of a prototypical $d_{x^2-y^2}$-wave altermagnet, one where the band structure alone gives altermagnetism \cite{Smejkal22a}, and another where electronic interactions along with the band structure generates altermagnetism \cite{Maier23}. 
We uncover that the characteristic spin-sublattice locking in both of these models put severe constraints on the nature of the superconducting pairing in altermagnets. 
In particular, we establish that the most common form of superconductivity, conventional or spin-singlet uniform, or onsite, $s$-wave pairing is not feasible in altermagnets. 
We further show that nearest-neighbor equal-spin-triplet $p$-wave pairing is also not possible to achieve. In fact, considering pairing up to next-nearest-neighbors in real space, our analysis
analysis reveals that the only allowed superconducting pairing symmetries in a $d_{x^2-y^2}$-wave altermagnet are spin-singlet $d_{x^2-y^2}$-wave and extended $s$-wave pairing, and mixed spin-triplet $p$-wave pairing. Longer-range pairing may also be allowed, but results most often in more nodes, resulting in a reduced superconducting condensation energy.
The only exception to these constraints is if we also allow for odd-frequency pairing \cite{Bergeret05,Linder19}, then uniform $s$-wave pairing becomes allowed in altermagnets as an spin-triplet odd-frequency superconducting state. Similarly, for odd-frequency superconductivity also spin-singlet $p$-wave, equal-spin-triplet onsite $s$- and $d_{xy}$-wave, and mixed spin-triplet $d_{x^2-y^2}$- and extended $s$-wave pairing symmetries become  allowed. These results are compactly summarized in Tables \ref{tab:table1} and \ref{tab:table2}.

In this work, we only consider the case of a $d_{x^2-y^2}$-wave altermagnet on a square lattice with a single orbital. However, our analysis can be easily extended to other forms of altermagnets on different lattices and also for multi-orbital systems. The key is to carefully consider the characteristic spin-sublattice coupling of altermagnets, which will generate constraints on the superconducting pairing. Moreover, the analysis of this work is very general and does not depend on whether superconductivity is intrinsic to the material or proximity induced. Hence, our findings are applicable both to systems exhibiting coexistence of superconductivity and altermagnetism, and to hybrid structures, such as  altermagnet-superconductor junctions. Thus the results in this work are important to a large class of systems.

\begin{acknowledgments}

We gratefully acknowledge financial support from the Knut and Alice Wallenberg Foundation through the Wallenberg Academy Fellows program, KAW 2019.0309, and the Swedish Research Council (Vetenskapsr\aa det) grant agreement no.~2022-03963. The computations were enabled by resources provided by the National Academic Infrastructure for Supercomputing in Sweden (NAISS), partially funded by the Swedish Research Council through grant agreement no.~2022-06725.

\end{acknowledgments}

 \bibliographystyle{apsrev4-1}
\bibliography{Cuprates}

\begin{thebibliography}{83}%
\makeatletter
\providecommand \@ifxundefined [1]{%
 \@ifx{#1\undefined}
}%
\providecommand \@ifnum [1]{%
 \ifnum #1\expandafter \@firstoftwo
 \else \expandafter \@secondoftwo
 \fi
}%
\providecommand \@ifx [1]{%
 \ifx #1\expandafter \@firstoftwo
 \else \expandafter \@secondoftwo
 \fi
}%
\providecommand \natexlab [1]{#1}%
\providecommand \enquote  [1]{``#1''}%
\providecommand \bibnamefont  [1]{#1}%
\providecommand \bibfnamefont [1]{#1}%
\providecommand \citenamefont [1]{#1}%
\providecommand \href@noop [0]{\@secondoftwo}%
\providecommand \href [0]{\begingroup \@sanitize@url \@href}%
\providecommand \@href[1]{\@@startlink{#1}\@@href}%
\providecommand \@@href[1]{\endgroup#1\@@endlink}%
\providecommand \@sanitize@url [0]{\catcode `\\12\catcode `\$12\catcode
  `\&12\catcode `\#12\catcode `\^12\catcode `\_12\catcode `\%12\relax}%
\providecommand \@@startlink[1]{}%
\providecommand \@@endlink[0]{}%
\providecommand \url  [0]{\begingroup\@sanitize@url \@url }%
\providecommand \@url [1]{\endgroup\@href {#1}{\urlprefix }}%
\providecommand \urlprefix  [0]{URL }%
\providecommand \Eprint [0]{\href }%
\providecommand \doibase [0]{http://dx.doi.org/}%
\providecommand \selectlanguage [0]{\@gobble}%
\providecommand \bibinfo  [0]{\@secondoftwo}%
\providecommand \bibfield  [0]{\@secondoftwo}%
\providecommand \translation [1]{[#1]}%
\providecommand \BibitemOpen [0]{}%
\providecommand \bibitemStop [0]{}%
\providecommand \bibitemNoStop [0]{.\EOS\space}%
\providecommand \EOS [0]{\spacefactor3000\relax}%
\providecommand \BibitemShut  [1]{\csname bibitem#1\endcsname}%
\let\auto@bib@innerbib\@empty
\bibitem [{\citenamefont {Scalapino}(1995)}]{Scalapino95}%
  \BibitemOpen
  \bibfield  {author} {\bibinfo {author} {\bibfnamefont {D.}~\bibnamefont
  {Scalapino}},\ }\href {\doibase https://doi.org/10.1016/0370-1573(94)00086-I}
  {\bibfield  {journal} {\bibinfo  {journal} {Physics Reports}\ }\textbf
  {\bibinfo {volume} {250}},\ \bibinfo {pages} {329 } (\bibinfo {year}
  {1995})}\BibitemShut {NoStop}%
\bibitem [{\citenamefont {Hirschfeld}\ \emph {et~al.}(2011)\citenamefont
  {Hirschfeld}, \citenamefont {Korshunov},\ and\ \citenamefont
  {Mazin}}]{Hirschfeld11}%
  \BibitemOpen
  \bibfield  {author} {\bibinfo {author} {\bibfnamefont {P.~J.}\ \bibnamefont
  {Hirschfeld}}, \bibinfo {author} {\bibfnamefont {M.~M.}\ \bibnamefont
  {Korshunov}}, \ and\ \bibinfo {author} {\bibfnamefont {I.~I.}\ \bibnamefont
  {Mazin}},\ }\href {\doibase 10.1088/0034-4885/74/12/124508} {\bibfield
  {journal} {\bibinfo  {journal} {Rep. Prog. Phys.}\ }\textbf {\bibinfo
  {volume} {74}},\ \bibinfo {pages} {124508} (\bibinfo {year}
  {2011})}\BibitemShut {NoStop}%
\bibitem [{\citenamefont {Chubukov}\ \emph {et~al.}(2008)\citenamefont
  {Chubukov}, \citenamefont {Pines},\ and\ \citenamefont {Schmalian}}]{sfbook}%
  \BibitemOpen
  \bibfield  {author} {\bibinfo {author} {\bibfnamefont {A.}~\bibnamefont
  {Chubukov}}, \bibinfo {author} {\bibfnamefont {D.}~\bibnamefont {Pines}}, \
  and\ \bibinfo {author} {\bibfnamefont {J.}~\bibnamefont {Schmalian}},\ }in\
  \href {\doibase 10.1007/978-3-540-73253-2_22} {\emph {\bibinfo {booktitle}
  {Superconductivity}}},\ \bibinfo {editor} {edited by\ \bibinfo {editor}
  {\bibfnamefont {K.}~\bibnamefont {Bennemann}}\ and\ \bibinfo {editor}
  {\bibfnamefont {J.}~\bibnamefont {Ketterson}}}\ (\bibinfo  {publisher}
  {Springer Berlin Heidelberg},\ \bibinfo {year} {2008})\BibitemShut {NoStop}%
\bibitem [{\citenamefont {Fulde}\ and\ \citenamefont
  {Ferrell}(1964)}]{Fulde64}%
  \BibitemOpen
  \bibfield  {author} {\bibinfo {author} {\bibfnamefont {P.}~\bibnamefont
  {Fulde}}\ and\ \bibinfo {author} {\bibfnamefont {R.~A.}\ \bibnamefont
  {Ferrell}},\ }\href {\doibase 10.1103/PhysRev.135.A550} {\bibfield  {journal}
  {\bibinfo  {journal} {Phys. Rev.}\ }\textbf {\bibinfo {volume} {135}},\
  \bibinfo {pages} {A550} (\bibinfo {year} {1964})}\BibitemShut {NoStop}%
\bibitem [{\citenamefont {Larkin}\ and\ \citenamefont
  {Ovchinnikov}(1964)}]{Larkin64}%
  \BibitemOpen
  \bibfield  {author} {\bibinfo {author} {\bibfnamefont {A.~I.}\ \bibnamefont
  {Larkin}}\ and\ \bibinfo {author} {\bibfnamefont {Y.~N.}\ \bibnamefont
  {Ovchinnikov}},\ }\href@noop {} {\bibfield  {journal} {\bibinfo  {journal}
  {Zh. Eksp. Teor. Fiz.}\ }\textbf {\bibinfo {volume} {47}},\ \bibinfo {pages}
  {1136} (\bibinfo {year} {1964})}\BibitemShut {NoStop}%
\bibitem [{\citenamefont {Alicea}(2012)}]{Alicea12}%
  \BibitemOpen
  \bibfield  {author} {\bibinfo {author} {\bibfnamefont {J.}~\bibnamefont
  {Alicea}},\ }\href {\doibase 10.1088/0034-4885/75/7/076501} {\bibfield
  {journal} {\bibinfo  {journal} {Rep. Prog. Phys.}\ }\textbf {\bibinfo
  {volume} {75}},\ \bibinfo {pages} {076501} (\bibinfo {year}
  {2012})}\BibitemShut {NoStop}%
\bibitem [{\citenamefont {Beenakker}(2013)}]{Beenakker13}%
  \BibitemOpen
  \bibfield  {author} {\bibinfo {author} {\bibfnamefont {C.}~\bibnamefont
  {Beenakker}},\ }\href {\doibase
  https://doi.org/10.1146/annurev-conmatphys-030212-184337} {\bibfield
  {journal} {\bibinfo  {journal} {Annu. Rev. Condens.}\ }\textbf {\bibinfo
  {volume} {4}},\ \bibinfo {pages} {113} (\bibinfo {year} {2013})}\BibitemShut
  {NoStop}%
\bibitem [{\citenamefont {Aguado}(2017)}]{Aguado17}%
  \BibitemOpen
  \bibfield  {author} {\bibinfo {author} {\bibfnamefont {R.}~\bibnamefont
  {Aguado}},\ }\href {\doibase 10.1393/ncr/i2017-10141-9} {\bibfield  {journal}
  {\bibinfo  {journal} {La Rivista del Nuovo Cimento}\ }\textbf {\bibinfo
  {volume} {40}},\ \bibinfo {pages} {523} (\bibinfo {year} {2017})}\BibitemShut
  {NoStop}%
\bibitem [{\citenamefont {Lutchyn}\ \emph {et~al.}(2018)\citenamefont
  {Lutchyn}, \citenamefont {Bakkers}, \citenamefont {Kouwenhoven},
  \citenamefont {Krogstrup}, \citenamefont {Marcus},\ and\ \citenamefont
  {Oreg}}]{Lutchyn18}%
  \BibitemOpen
  \bibfield  {author} {\bibinfo {author} {\bibfnamefont {R.~M.}\ \bibnamefont
  {Lutchyn}}, \bibinfo {author} {\bibfnamefont {E.~P. A.~M.}\ \bibnamefont
  {Bakkers}}, \bibinfo {author} {\bibfnamefont {L.~P.}\ \bibnamefont
  {Kouwenhoven}}, \bibinfo {author} {\bibfnamefont {P.}~\bibnamefont
  {Krogstrup}}, \bibinfo {author} {\bibfnamefont {C.~M.}\ \bibnamefont
  {Marcus}}, \ and\ \bibinfo {author} {\bibfnamefont {Y.}~\bibnamefont
  {Oreg}},\ }\href {\doibase 10.1038/s41578-018-0003-1} {\bibfield  {journal}
  {\bibinfo  {journal} {Nat. Rev. Mater.}\ }\textbf {\bibinfo {volume} {3}},\
  \bibinfo {pages} {52} (\bibinfo {year} {2018})}\BibitemShut {NoStop}%
\bibitem [{\citenamefont {Flensberg}\ \emph {et~al.}(2021)\citenamefont
  {Flensberg}, \citenamefont {von Oppen},\ and\ \citenamefont
  {Stern}}]{Flensberg21}%
  \BibitemOpen
  \bibfield  {author} {\bibinfo {author} {\bibfnamefont {K.}~\bibnamefont
  {Flensberg}}, \bibinfo {author} {\bibfnamefont {F.}~\bibnamefont {von
  Oppen}}, \ and\ \bibinfo {author} {\bibfnamefont {A.}~\bibnamefont {Stern}},\
  }\href {\doibase 10.1038/s41578-021-00336-6} {\bibfield  {journal} {\bibinfo
  {journal} {Nat. Rev. Mater.}\ }\textbf {\bibinfo {volume} {6}},\ \bibinfo
  {pages} {944} (\bibinfo {year} {2021})}\BibitemShut {NoStop}%
\bibitem [{\citenamefont {Eschrig}\ and\ \citenamefont
  {L{\"o}fwander}(2008)}]{Eschrig08}%
  \BibitemOpen
  \bibfield  {author} {\bibinfo {author} {\bibfnamefont {M.}~\bibnamefont
  {Eschrig}}\ and\ \bibinfo {author} {\bibfnamefont {T.}~\bibnamefont
  {L{\"o}fwander}},\ }\href {\doibase 10.1038/nphys831} {\bibfield  {journal}
  {\bibinfo  {journal} {Nat. Phys.}\ }\textbf {\bibinfo {volume} {4}},\
  \bibinfo {pages} {138} (\bibinfo {year} {2008})}\BibitemShut {NoStop}%
\bibitem [{\citenamefont {Linder}\ and\ \citenamefont
  {Robinson}(2015)}]{Linder15}%
  \BibitemOpen
  \bibfield  {author} {\bibinfo {author} {\bibfnamefont {J.}~\bibnamefont
  {Linder}}\ and\ \bibinfo {author} {\bibfnamefont {J.~W.~A.}\ \bibnamefont
  {Robinson}},\ }\href {\doibase 10.1038/nphys3242} {\bibfield  {journal}
  {\bibinfo  {journal} {Nat. Phys.}\ }\textbf {\bibinfo {volume} {11}},\
  \bibinfo {pages} {307} (\bibinfo {year} {2015})}\BibitemShut {NoStop}%
\bibitem [{\citenamefont {Šmejkal}\ \emph {et~al.}(2020)\citenamefont
  {Šmejkal}, \citenamefont {González-Hernández}, \citenamefont {Jungwirth},\
  and\ \citenamefont {Sinova}}]{Smejkal20}%
  \BibitemOpen
  \bibfield  {author} {\bibinfo {author} {\bibfnamefont {L.}~\bibnamefont
  {Šmejkal}}, \bibinfo {author} {\bibfnamefont {R.}~\bibnamefont
  {González-Hernández}}, \bibinfo {author} {\bibfnamefont {T.}~\bibnamefont
  {Jungwirth}}, \ and\ \bibinfo {author} {\bibfnamefont {J.}~\bibnamefont
  {Sinova}},\ }\href {\doibase 10.1126/sciadv.aaz8809} {\bibfield  {journal}
  {\bibinfo  {journal} {Sci. Adv.}\ }\textbf {\bibinfo {volume} {6}},\ \bibinfo
  {pages} {eaaz8809} (\bibinfo {year} {2020})}\BibitemShut {NoStop}%
\bibitem [{\citenamefont {Yuan}\ \emph {et~al.}(2020)\citenamefont {Yuan},
  \citenamefont {Wang}, \citenamefont {Luo}, \citenamefont {Rashba},\ and\
  \citenamefont {Zunger}}]{Yuan20}%
  \BibitemOpen
  \bibfield  {author} {\bibinfo {author} {\bibfnamefont {L.-D.}\ \bibnamefont
  {Yuan}}, \bibinfo {author} {\bibfnamefont {Z.}~\bibnamefont {Wang}}, \bibinfo
  {author} {\bibfnamefont {J.-W.}\ \bibnamefont {Luo}}, \bibinfo {author}
  {\bibfnamefont {E.~I.}\ \bibnamefont {Rashba}}, \ and\ \bibinfo {author}
  {\bibfnamefont {A.}~\bibnamefont {Zunger}},\ }\href {\doibase
  10.1103/PhysRevB.102.014422} {\bibfield  {journal} {\bibinfo  {journal}
  {Phys. Rev. B}\ }\textbf {\bibinfo {volume} {102}},\ \bibinfo {pages}
  {014422} (\bibinfo {year} {2020})}\BibitemShut {NoStop}%
\bibitem [{\citenamefont {Mazin}\ \emph {et~al.}(2021)\citenamefont {Mazin},
  \citenamefont {Koepernik}, \citenamefont {Johannes}, \citenamefont
  {González-Hernández},\ and\ \citenamefont {Šmejkal}}]{Mazin21}%
  \BibitemOpen
  \bibfield  {author} {\bibinfo {author} {\bibfnamefont {I.~I.}\ \bibnamefont
  {Mazin}}, \bibinfo {author} {\bibfnamefont {K.}~\bibnamefont {Koepernik}},
  \bibinfo {author} {\bibfnamefont {M.~D.}\ \bibnamefont {Johannes}}, \bibinfo
  {author} {\bibfnamefont {R.}~\bibnamefont {González-Hernández}}, \ and\
  \bibinfo {author} {\bibfnamefont {L.}~\bibnamefont {Šmejkal}},\ }\href
  {\doibase 10.1073/pnas.2108924118} {\bibfield  {journal} {\bibinfo  {journal}
  {Proc. Natl. Acad. Sci.}\ }\textbf {\bibinfo {volume} {118}},\ \bibinfo
  {pages} {e2108924118} (\bibinfo {year} {2021})}\BibitemShut {NoStop}%
\bibitem [{\citenamefont {\ifmmode~\check{S}\else \v{S}\fi{}mejkal}\ \emph
  {et~al.}(2022{\natexlab{a}})\citenamefont {\ifmmode~\check{S}\else
  \v{S}\fi{}mejkal}, \citenamefont {Sinova},\ and\ \citenamefont
  {Jungwirth}}]{Smejkal22a}%
  \BibitemOpen
  \bibfield  {author} {\bibinfo {author} {\bibfnamefont {L.}~\bibnamefont
  {\ifmmode~\check{S}\else \v{S}\fi{}mejkal}}, \bibinfo {author} {\bibfnamefont
  {J.}~\bibnamefont {Sinova}}, \ and\ \bibinfo {author} {\bibfnamefont
  {T.}~\bibnamefont {Jungwirth}},\ }\href {\doibase 10.1103/PhysRevX.12.031042}
  {\bibfield  {journal} {\bibinfo  {journal} {Phys. Rev. X}\ }\textbf {\bibinfo
  {volume} {12}},\ \bibinfo {pages} {031042} (\bibinfo {year}
  {2022}{\natexlab{a}})}\BibitemShut {NoStop}%
\bibitem [{\citenamefont {\ifmmode~\check{S}\else \v{S}\fi{}mejkal}\ \emph
  {et~al.}(2022{\natexlab{b}})\citenamefont {\ifmmode~\check{S}\else
  \v{S}\fi{}mejkal}, \citenamefont {Sinova},\ and\ \citenamefont
  {Jungwirth}}]{Smejkal22}%
  \BibitemOpen
  \bibfield  {author} {\bibinfo {author} {\bibfnamefont {L.}~\bibnamefont
  {\ifmmode~\check{S}\else \v{S}\fi{}mejkal}}, \bibinfo {author} {\bibfnamefont
  {J.}~\bibnamefont {Sinova}}, \ and\ \bibinfo {author} {\bibfnamefont
  {T.}~\bibnamefont {Jungwirth}},\ }\href {\doibase 10.1103/PhysRevX.12.040501}
  {\bibfield  {journal} {\bibinfo  {journal} {Phys. Rev. X}\ }\textbf {\bibinfo
  {volume} {12}},\ \bibinfo {pages} {040501} (\bibinfo {year}
  {2022}{\natexlab{b}})}\BibitemShut {NoStop}%
\bibitem [{\citenamefont {Hayami}\ \emph {et~al.}(2019)\citenamefont {Hayami},
  \citenamefont {Yanagi},\ and\ \citenamefont {Kusunose}}]{Hayami19}%
  \BibitemOpen
  \bibfield  {author} {\bibinfo {author} {\bibfnamefont {S.}~\bibnamefont
  {Hayami}}, \bibinfo {author} {\bibfnamefont {Y.}~\bibnamefont {Yanagi}}, \
  and\ \bibinfo {author} {\bibfnamefont {H.}~\bibnamefont {Kusunose}},\ }\href
  {http://dx.doi.org/10.7566/JPSJ.88.123702} {\bibfield  {journal} {\bibinfo
  {journal} {J. Phys. Soc. Jpn.}\ }\textbf {\bibinfo {volume} {88}},\ \bibinfo
  {pages} {123702} (\bibinfo {year} {2019})}\BibitemShut {NoStop}%
\bibitem [{\citenamefont {Hayami}\ \emph {et~al.}(2020)\citenamefont {Hayami},
  \citenamefont {Yanagi},\ and\ \citenamefont {Kusunose}}]{Hayami20}%
  \BibitemOpen
  \bibfield  {author} {\bibinfo {author} {\bibfnamefont {S.}~\bibnamefont
  {Hayami}}, \bibinfo {author} {\bibfnamefont {Y.}~\bibnamefont {Yanagi}}, \
  and\ \bibinfo {author} {\bibfnamefont {H.}~\bibnamefont {Kusunose}},\ }\href
  {\doibase 10.1103/PhysRevB.102.144441} {\bibfield  {journal} {\bibinfo
  {journal} {Phys. Rev. B}\ }\textbf {\bibinfo {volume} {102}},\ \bibinfo
  {pages} {144441} (\bibinfo {year} {2020})}\BibitemShut {NoStop}%
\bibitem [{\citenamefont {Feng}\ \emph {et~al.}(2022)\citenamefont {Feng},
  \citenamefont {Zhou}, \citenamefont {{\v{S}}mejkal}, \citenamefont {Wu},
  \citenamefont {Zhu}, \citenamefont {Guo}, \citenamefont
  {Gonz{\'a}lez-Hern{\'a}ndez}, \citenamefont {Wang}, \citenamefont {Yan},
  \citenamefont {Qin}, \citenamefont {Zhang}, \citenamefont {Wu}, \citenamefont
  {Chen}, \citenamefont {Meng}, \citenamefont {Liu}, \citenamefont {Xia},
  \citenamefont {Sinova}, \citenamefont {Jungwirth},\ and\ \citenamefont
  {Liu}}]{Feng22}%
  \BibitemOpen
  \bibfield  {author} {\bibinfo {author} {\bibfnamefont {Z.}~\bibnamefont
  {Feng}}, \bibinfo {author} {\bibfnamefont {X.}~\bibnamefont {Zhou}}, \bibinfo
  {author} {\bibfnamefont {L.}~\bibnamefont {{\v{S}}mejkal}}, \bibinfo {author}
  {\bibfnamefont {L.}~\bibnamefont {Wu}}, \bibinfo {author} {\bibfnamefont
  {Z.}~\bibnamefont {Zhu}}, \bibinfo {author} {\bibfnamefont {H.}~\bibnamefont
  {Guo}}, \bibinfo {author} {\bibfnamefont {R.}~\bibnamefont
  {Gonz{\'a}lez-Hern{\'a}ndez}}, \bibinfo {author} {\bibfnamefont
  {X.}~\bibnamefont {Wang}}, \bibinfo {author} {\bibfnamefont {H.}~\bibnamefont
  {Yan}}, \bibinfo {author} {\bibfnamefont {P.}~\bibnamefont {Qin}}, \bibinfo
  {author} {\bibfnamefont {X.}~\bibnamefont {Zhang}}, \bibinfo {author}
  {\bibfnamefont {H.}~\bibnamefont {Wu}}, \bibinfo {author} {\bibfnamefont
  {H.}~\bibnamefont {Chen}}, \bibinfo {author} {\bibfnamefont {Z.}~\bibnamefont
  {Meng}}, \bibinfo {author} {\bibfnamefont {L.}~\bibnamefont {Liu}}, \bibinfo
  {author} {\bibfnamefont {Z.}~\bibnamefont {Xia}}, \bibinfo {author}
  {\bibfnamefont {J.}~\bibnamefont {Sinova}}, \bibinfo {author} {\bibfnamefont
  {T.}~\bibnamefont {Jungwirth}}, \ and\ \bibinfo {author} {\bibfnamefont
  {Z.}~\bibnamefont {Liu}},\ }\href {\doibase 10.1038/s41928-022-00866-z}
  {\bibfield  {journal} {\bibinfo  {journal} {Nat. Electron.}\ }\textbf
  {\bibinfo {volume} {5}},\ \bibinfo {pages} {735} (\bibinfo {year}
  {2022})}\BibitemShut {NoStop}%
\bibitem [{\citenamefont {Gonzalez~Betancourt}\ \emph
  {et~al.}(2023)\citenamefont {Gonzalez~Betancourt}, \citenamefont
  {Zub\'a\ifmmode~\check{c}\else \v{c}\fi{}}, \citenamefont
  {Gonzalez-Hernandez}, \citenamefont {Geishendorf}, \citenamefont {\ifmmode
  \check{S}\else \v{S}\fi{}ob\'a\ifmmode~\check{n}\else \v{n}\fi{}},
  \citenamefont {Springholz}, \citenamefont {Olejn\'{\i}k}, \citenamefont
  {\ifmmode~\check{S}\else \v{S}\fi{}mejkal}, \citenamefont {Sinova},
  \citenamefont {Jungwirth}, \citenamefont {Goennenwein}, \citenamefont
  {Thomas}, \citenamefont {Reichlov\'a}, \citenamefont {\ifmmode~\check{Z}\else
  \v{Z}\fi{}elezn\'y},\ and\ \citenamefont {Kriegner}}]{Gonzalez23}%
  \BibitemOpen
  \bibfield  {author} {\bibinfo {author} {\bibfnamefont {R.~D.}\ \bibnamefont
  {Gonzalez~Betancourt}}, \bibinfo {author} {\bibfnamefont {J.}~\bibnamefont
  {Zub\'a\ifmmode~\check{c}\else \v{c}\fi{}}}, \bibinfo {author} {\bibfnamefont
  {R.}~\bibnamefont {Gonzalez-Hernandez}}, \bibinfo {author} {\bibfnamefont
  {K.}~\bibnamefont {Geishendorf}}, \bibinfo {author} {\bibfnamefont
  {Z.}~\bibnamefont {\ifmmode \check{S}\else
  \v{S}\fi{}ob\'a\ifmmode~\check{n}\else \v{n}\fi{}}}, \bibinfo {author}
  {\bibfnamefont {G.}~\bibnamefont {Springholz}}, \bibinfo {author}
  {\bibfnamefont {K.}~\bibnamefont {Olejn\'{\i}k}}, \bibinfo {author}
  {\bibfnamefont {L.}~\bibnamefont {\ifmmode~\check{S}\else \v{S}\fi{}mejkal}},
  \bibinfo {author} {\bibfnamefont {J.}~\bibnamefont {Sinova}}, \bibinfo
  {author} {\bibfnamefont {T.}~\bibnamefont {Jungwirth}}, \bibinfo {author}
  {\bibfnamefont {S.~T.~B.}\ \bibnamefont {Goennenwein}}, \bibinfo {author}
  {\bibfnamefont {A.}~\bibnamefont {Thomas}}, \bibinfo {author} {\bibfnamefont
  {H.}~\bibnamefont {Reichlov\'a}}, \bibinfo {author} {\bibfnamefont
  {J.}~\bibnamefont {\ifmmode~\check{Z}\else \v{Z}\fi{}elezn\'y}}, \ and\
  \bibinfo {author} {\bibfnamefont {D.}~\bibnamefont {Kriegner}},\ }\href
  {\doibase 10.1103/PhysRevLett.130.036702} {\bibfield  {journal} {\bibinfo
  {journal} {Phys. Rev. Lett.}\ }\textbf {\bibinfo {volume} {130}},\ \bibinfo
  {pages} {036702} (\bibinfo {year} {2023})}\BibitemShut {NoStop}%
\bibitem [{\citenamefont {Bai}\ \emph {et~al.}(2023)\citenamefont {Bai},
  \citenamefont {Zhang}, \citenamefont {Zhou}, \citenamefont {Chen},
  \citenamefont {Wan}, \citenamefont {Han}, \citenamefont {Zhu}, \citenamefont
  {Liang}, \citenamefont {Su}, \citenamefont {Han}, \citenamefont {Pan},\ and\
  \citenamefont {Song}}]{Bai23}%
  \BibitemOpen
  \bibfield  {author} {\bibinfo {author} {\bibfnamefont {H.}~\bibnamefont
  {Bai}}, \bibinfo {author} {\bibfnamefont {Y.~C.}\ \bibnamefont {Zhang}},
  \bibinfo {author} {\bibfnamefont {Y.~J.}\ \bibnamefont {Zhou}}, \bibinfo
  {author} {\bibfnamefont {P.}~\bibnamefont {Chen}}, \bibinfo {author}
  {\bibfnamefont {C.~H.}\ \bibnamefont {Wan}}, \bibinfo {author} {\bibfnamefont
  {L.}~\bibnamefont {Han}}, \bibinfo {author} {\bibfnamefont {W.~X.}\
  \bibnamefont {Zhu}}, \bibinfo {author} {\bibfnamefont {S.~X.}\ \bibnamefont
  {Liang}}, \bibinfo {author} {\bibfnamefont {Y.~C.}\ \bibnamefont {Su}},
  \bibinfo {author} {\bibfnamefont {X.~F.}\ \bibnamefont {Han}}, \bibinfo
  {author} {\bibfnamefont {F.}~\bibnamefont {Pan}}, \ and\ \bibinfo {author}
  {\bibfnamefont {C.}~\bibnamefont {Song}},\ }\href {\doibase
  10.1103/PhysRevLett.130.216701} {\bibfield  {journal} {\bibinfo  {journal}
  {Phys. Rev. Lett.}\ }\textbf {\bibinfo {volume} {130}},\ \bibinfo {pages}
  {216701} (\bibinfo {year} {2023})}\BibitemShut {NoStop}%
\bibitem [{\citenamefont {Fedchenko}\ \emph {et~al.}(2024)\citenamefont
  {Fedchenko}, \citenamefont {Minár}, \citenamefont {Akashdeep}, \citenamefont
  {D’Souza}, \citenamefont {Vasilyev}, \citenamefont {Tkach}, \citenamefont
  {Odenbreit}, \citenamefont {Nguyen}, \citenamefont {Kutnyakhov},
  \citenamefont {Wind}, \citenamefont {Wenthaus}, \citenamefont {Scholz},
  \citenamefont {Rossnagel}, \citenamefont {Hoesch}, \citenamefont
  {Aeschlimann}, \citenamefont {Stadtmüller}, \citenamefont {Kläui},
  \citenamefont {Schönhense}, \citenamefont {Jungwirth}, \citenamefont
  {Hellenes}, \citenamefont {Jakob}, \citenamefont {Šmejkal}, \citenamefont
  {Sinova},\ and\ \citenamefont {Elmers}}]{Olena24}%
  \BibitemOpen
  \bibfield  {author} {\bibinfo {author} {\bibfnamefont {O.}~\bibnamefont
  {Fedchenko}}, \bibinfo {author} {\bibfnamefont {J.}~\bibnamefont {Minár}},
  \bibinfo {author} {\bibfnamefont {A.}~\bibnamefont {Akashdeep}}, \bibinfo
  {author} {\bibfnamefont {S.~W.}\ \bibnamefont {D’Souza}}, \bibinfo {author}
  {\bibfnamefont {D.}~\bibnamefont {Vasilyev}}, \bibinfo {author}
  {\bibfnamefont {O.}~\bibnamefont {Tkach}}, \bibinfo {author} {\bibfnamefont
  {L.}~\bibnamefont {Odenbreit}}, \bibinfo {author} {\bibfnamefont
  {Q.}~\bibnamefont {Nguyen}}, \bibinfo {author} {\bibfnamefont
  {D.}~\bibnamefont {Kutnyakhov}}, \bibinfo {author} {\bibfnamefont
  {N.}~\bibnamefont {Wind}}, \bibinfo {author} {\bibfnamefont {L.}~\bibnamefont
  {Wenthaus}}, \bibinfo {author} {\bibfnamefont {M.}~\bibnamefont {Scholz}},
  \bibinfo {author} {\bibfnamefont {K.}~\bibnamefont {Rossnagel}}, \bibinfo
  {author} {\bibfnamefont {M.}~\bibnamefont {Hoesch}}, \bibinfo {author}
  {\bibfnamefont {M.}~\bibnamefont {Aeschlimann}}, \bibinfo {author}
  {\bibfnamefont {B.}~\bibnamefont {Stadtmüller}}, \bibinfo {author}
  {\bibfnamefont {M.}~\bibnamefont {Kläui}}, \bibinfo {author} {\bibfnamefont
  {G.}~\bibnamefont {Schönhense}}, \bibinfo {author} {\bibfnamefont
  {T.}~\bibnamefont {Jungwirth}}, \bibinfo {author} {\bibfnamefont {A.~B.}\
  \bibnamefont {Hellenes}}, \bibinfo {author} {\bibfnamefont {G.}~\bibnamefont
  {Jakob}}, \bibinfo {author} {\bibfnamefont {L.}~\bibnamefont {Šmejkal}},
  \bibinfo {author} {\bibfnamefont {J.}~\bibnamefont {Sinova}}, \ and\ \bibinfo
  {author} {\bibfnamefont {H.-J.}\ \bibnamefont {Elmers}},\ }\href {\doibase
  10.1126/sciadv.adj4883} {\bibfield  {journal} {\bibinfo  {journal} {Sci.
  Adv.}\ }\textbf {\bibinfo {volume} {10}},\ \bibinfo {pages} {eadj4883}
  (\bibinfo {year} {2024})}\BibitemShut {NoStop}%
\bibitem [{\citenamefont {Zhu}\ \emph {et~al.}(2024)\citenamefont {Zhu},
  \citenamefont {Chen}, \citenamefont {Liu}, \citenamefont {Liu}, \citenamefont
  {Liu}, \citenamefont {Zha}, \citenamefont {Qu}, \citenamefont {Hong},
  \citenamefont {Li}, \citenamefont {Jiang}, \citenamefont {Ma}, \citenamefont
  {Hao}, \citenamefont {Zhu}, \citenamefont {Liu}, \citenamefont {Zeng},
  \citenamefont {Jayaram}, \citenamefont {Lenger}, \citenamefont {Ding},
  \citenamefont {Mo}, \citenamefont {Tanaka}, \citenamefont {Arita},
  \citenamefont {Liu}, \citenamefont {Ye}, \citenamefont {Shen}, \citenamefont
  {Wrachtrup}, \citenamefont {Huang}, \citenamefont {He}, \citenamefont {Qiao},
  \citenamefont {Liu},\ and\ \citenamefont {Liu}}]{Zhu24}%
  \BibitemOpen
  \bibfield  {author} {\bibinfo {author} {\bibfnamefont {Y.-P.}\ \bibnamefont
  {Zhu}}, \bibinfo {author} {\bibfnamefont {X.}~\bibnamefont {Chen}}, \bibinfo
  {author} {\bibfnamefont {X.-R.}\ \bibnamefont {Liu}}, \bibinfo {author}
  {\bibfnamefont {Y.}~\bibnamefont {Liu}}, \bibinfo {author} {\bibfnamefont
  {P.}~\bibnamefont {Liu}}, \bibinfo {author} {\bibfnamefont {H.}~\bibnamefont
  {Zha}}, \bibinfo {author} {\bibfnamefont {G.}~\bibnamefont {Qu}}, \bibinfo
  {author} {\bibfnamefont {C.}~\bibnamefont {Hong}}, \bibinfo {author}
  {\bibfnamefont {J.}~\bibnamefont {Li}}, \bibinfo {author} {\bibfnamefont
  {Z.}~\bibnamefont {Jiang}}, \bibinfo {author} {\bibfnamefont {X.-M.}\
  \bibnamefont {Ma}}, \bibinfo {author} {\bibfnamefont {Y.-J.}\ \bibnamefont
  {Hao}}, \bibinfo {author} {\bibfnamefont {M.-Y.}\ \bibnamefont {Zhu}},
  \bibinfo {author} {\bibfnamefont {W.}~\bibnamefont {Liu}}, \bibinfo {author}
  {\bibfnamefont {M.}~\bibnamefont {Zeng}}, \bibinfo {author} {\bibfnamefont
  {S.}~\bibnamefont {Jayaram}}, \bibinfo {author} {\bibfnamefont
  {M.}~\bibnamefont {Lenger}}, \bibinfo {author} {\bibfnamefont
  {J.}~\bibnamefont {Ding}}, \bibinfo {author} {\bibfnamefont {S.}~\bibnamefont
  {Mo}}, \bibinfo {author} {\bibfnamefont {K.}~\bibnamefont {Tanaka}}, \bibinfo
  {author} {\bibfnamefont {M.}~\bibnamefont {Arita}}, \bibinfo {author}
  {\bibfnamefont {Z.}~\bibnamefont {Liu}}, \bibinfo {author} {\bibfnamefont
  {M.}~\bibnamefont {Ye}}, \bibinfo {author} {\bibfnamefont {D.}~\bibnamefont
  {Shen}}, \bibinfo {author} {\bibfnamefont {J.}~\bibnamefont {Wrachtrup}},
  \bibinfo {author} {\bibfnamefont {Y.}~\bibnamefont {Huang}}, \bibinfo
  {author} {\bibfnamefont {R.-H.}\ \bibnamefont {He}}, \bibinfo {author}
  {\bibfnamefont {S.}~\bibnamefont {Qiao}}, \bibinfo {author} {\bibfnamefont
  {Q.}~\bibnamefont {Liu}}, \ and\ \bibinfo {author} {\bibfnamefont
  {C.}~\bibnamefont {Liu}},\ }\href {\doibase 10.1038/s41586-024-07023-w}
  {\bibfield  {journal} {\bibinfo  {journal} {Nature}\ }\textbf {\bibinfo
  {volume} {626}},\ \bibinfo {pages} {523} (\bibinfo {year}
  {2024})}\BibitemShut {NoStop}%
\bibitem [{\citenamefont {Krempask{\'y}}\ \emph {et~al.}(2024)\citenamefont
  {Krempask{\'y}}, \citenamefont {{\v{S}}mejkal}, \citenamefont {D'Souza},
  \citenamefont {Hajlaoui}, \citenamefont {Springholz}, \citenamefont
  {Uhl{\'i}{\v{r}}ov{\'a}}, \citenamefont {Alarab}, \citenamefont
  {Constantinou}, \citenamefont {Strocov}, \citenamefont {Usanov},
  \citenamefont {Pudelko}, \citenamefont {Gonz{\'a}lez-Hern{\'a}ndez},
  \citenamefont {Birk~Hellenes}, \citenamefont {Jansa}, \citenamefont
  {Reichlov{\'a}}, \citenamefont {{\v{S}}ob{\'a}{\v{n}}}, \citenamefont
  {Gonzalez~Betancourt}, \citenamefont {Wadley}, \citenamefont {Sinova},
  \citenamefont {Kriegner}, \citenamefont {Min{\'a}r}, \citenamefont {Dil},\
  and\ \citenamefont {Jungwirth}}]{Krempasky24}%
  \BibitemOpen
  \bibfield  {author} {\bibinfo {author} {\bibfnamefont {J.}~\bibnamefont
  {Krempask{\'y}}}, \bibinfo {author} {\bibfnamefont {L.}~\bibnamefont
  {{\v{S}}mejkal}}, \bibinfo {author} {\bibfnamefont {S.~W.}\ \bibnamefont
  {D'Souza}}, \bibinfo {author} {\bibfnamefont {M.}~\bibnamefont {Hajlaoui}},
  \bibinfo {author} {\bibfnamefont {G.}~\bibnamefont {Springholz}}, \bibinfo
  {author} {\bibfnamefont {K.}~\bibnamefont {Uhl{\'i}{\v{r}}ov{\'a}}}, \bibinfo
  {author} {\bibfnamefont {F.}~\bibnamefont {Alarab}}, \bibinfo {author}
  {\bibfnamefont {P.~C.}\ \bibnamefont {Constantinou}}, \bibinfo {author}
  {\bibfnamefont {V.}~\bibnamefont {Strocov}}, \bibinfo {author} {\bibfnamefont
  {D.}~\bibnamefont {Usanov}}, \bibinfo {author} {\bibfnamefont {W.~R.}\
  \bibnamefont {Pudelko}}, \bibinfo {author} {\bibfnamefont {R.}~\bibnamefont
  {Gonz{\'a}lez-Hern{\'a}ndez}}, \bibinfo {author} {\bibfnamefont
  {A.}~\bibnamefont {Birk~Hellenes}}, \bibinfo {author} {\bibfnamefont
  {Z.}~\bibnamefont {Jansa}}, \bibinfo {author} {\bibfnamefont
  {H.}~\bibnamefont {Reichlov{\'a}}}, \bibinfo {author} {\bibfnamefont
  {Z.}~\bibnamefont {{\v{S}}ob{\'a}{\v{n}}}}, \bibinfo {author} {\bibfnamefont
  {R.~D.}\ \bibnamefont {Gonzalez~Betancourt}}, \bibinfo {author}
  {\bibfnamefont {P.}~\bibnamefont {Wadley}}, \bibinfo {author} {\bibfnamefont
  {J.}~\bibnamefont {Sinova}}, \bibinfo {author} {\bibfnamefont
  {D.}~\bibnamefont {Kriegner}}, \bibinfo {author} {\bibfnamefont
  {J.}~\bibnamefont {Min{\'a}r}}, \bibinfo {author} {\bibfnamefont {J.~H.}\
  \bibnamefont {Dil}}, \ and\ \bibinfo {author} {\bibfnamefont
  {T.}~\bibnamefont {Jungwirth}},\ }\href {\doibase 10.1038/s41586-023-06907-7}
  {\bibfield  {journal} {\bibinfo  {journal} {Nature}\ }\textbf {\bibinfo
  {volume} {626}},\ \bibinfo {pages} {517} (\bibinfo {year}
  {2024})}\BibitemShut {NoStop}%
\bibitem [{\citenamefont {Bai}\ \emph {et~al.}(2024)\citenamefont {Bai},
  \citenamefont {Feng}, \citenamefont {Liu}, \citenamefont {Šmejkal},
  \citenamefont {Mokrousov},\ and\ \citenamefont {Yao}}]{bai24}%
  \BibitemOpen
  \bibfield  {author} {\bibinfo {author} {\bibfnamefont {L.}~\bibnamefont
  {Bai}}, \bibinfo {author} {\bibfnamefont {W.}~\bibnamefont {Feng}}, \bibinfo
  {author} {\bibfnamefont {S.}~\bibnamefont {Liu}}, \bibinfo {author}
  {\bibfnamefont {L.}~\bibnamefont {Šmejkal}}, \bibinfo {author}
  {\bibfnamefont {Y.}~\bibnamefont {Mokrousov}}, \ and\ \bibinfo {author}
  {\bibfnamefont {Y.}~\bibnamefont {Yao}},\ }\href
  {https://arxiv.org/abs/2406.02123} {\bibfield  {journal} {\bibinfo  {journal}
  {arXiv:2406.02123}\ } (\bibinfo {year} {2024})}\BibitemShut {NoStop}%
\bibitem [{\citenamefont {Mazin}(2024)}]{Mazin24}%
  \BibitemOpen
  \bibfield  {author} {\bibinfo {author} {\bibfnamefont {I.}~\bibnamefont
  {Mazin}},\ }\href {\doibase DOI:10.1103/Physics.17.4} {\bibfield  {journal}
  {\bibinfo  {journal} {Physics}\ }\textbf {\bibinfo {volume} {17}},\ \bibinfo
  {pages} {4} (\bibinfo {year} {2024})}\BibitemShut {NoStop}%
\bibitem [{\citenamefont {Maier}\ and\ \citenamefont
  {Okamoto}(2023)}]{Maier23}%
  \BibitemOpen
  \bibfield  {author} {\bibinfo {author} {\bibfnamefont {T.~A.}\ \bibnamefont
  {Maier}}\ and\ \bibinfo {author} {\bibfnamefont {S.}~\bibnamefont
  {Okamoto}},\ }\href {\doibase 10.1103/PhysRevB.108.L100402} {\bibfield
  {journal} {\bibinfo  {journal} {Phys. Rev. B}\ }\textbf {\bibinfo {volume}
  {108}},\ \bibinfo {pages} {L100402} (\bibinfo {year} {2023})}\BibitemShut
  {NoStop}%
\bibitem [{\citenamefont {Leeb}\ \emph {et~al.}(2024)\citenamefont {Leeb},
  \citenamefont {Mook}, \citenamefont {\ifmmode~\check{S}\else
  \v{S}\fi{}mejkal},\ and\ \citenamefont {Knolle}}]{Leeb24}%
  \BibitemOpen
  \bibfield  {author} {\bibinfo {author} {\bibfnamefont {V.}~\bibnamefont
  {Leeb}}, \bibinfo {author} {\bibfnamefont {A.}~\bibnamefont {Mook}}, \bibinfo
  {author} {\bibfnamefont {L.}~\bibnamefont {\ifmmode~\check{S}\else
  \v{S}\fi{}mejkal}}, \ and\ \bibinfo {author} {\bibfnamefont {J.}~\bibnamefont
  {Knolle}},\ }\href {\doibase 10.1103/PhysRevLett.132.236701} {\bibfield
  {journal} {\bibinfo  {journal} {Phys. Rev. Lett.}\ }\textbf {\bibinfo
  {volume} {132}},\ \bibinfo {pages} {236701} (\bibinfo {year}
  {2024})}\BibitemShut {NoStop}%
\bibitem [{\citenamefont {Brekke}\ \emph {et~al.}(2023)\citenamefont {Brekke},
  \citenamefont {Brataas},\ and\ \citenamefont {Sudb\o{}}}]{Brekke23}%
  \BibitemOpen
  \bibfield  {author} {\bibinfo {author} {\bibfnamefont {B.}~\bibnamefont
  {Brekke}}, \bibinfo {author} {\bibfnamefont {A.}~\bibnamefont {Brataas}}, \
  and\ \bibinfo {author} {\bibfnamefont {A.}~\bibnamefont {Sudb\o{}}},\ }\href
  {\doibase 10.1103/PhysRevB.108.224421} {\bibfield  {journal} {\bibinfo
  {journal} {Phys. Rev. B}\ }\textbf {\bibinfo {volume} {108}},\ \bibinfo
  {pages} {224421} (\bibinfo {year} {2023})}\BibitemShut {NoStop}%
\bibitem [{\citenamefont {Roig}\ \emph {et~al.}(2024)\citenamefont {Roig},
  \citenamefont {Kreisel}, \citenamefont {Yu}, \citenamefont {Andersen},\ and\
  \citenamefont {Agterberg}}]{Roig24}%
  \BibitemOpen
  \bibfield  {author} {\bibinfo {author} {\bibfnamefont {M.}~\bibnamefont
  {Roig}}, \bibinfo {author} {\bibfnamefont {A.}~\bibnamefont {Kreisel}},
  \bibinfo {author} {\bibfnamefont {Y.}~\bibnamefont {Yu}}, \bibinfo {author}
  {\bibfnamefont {B.~M.}\ \bibnamefont {Andersen}}, \ and\ \bibinfo {author}
  {\bibfnamefont {D.~F.}\ \bibnamefont {Agterberg}},\ }\href
  {https://arxiv.org/abs/2402.15616} {\bibfield  {journal} {\bibinfo  {journal}
  {arXiv:2402.15616}\ } (\bibinfo {year} {2024})}\BibitemShut {NoStop}%
\bibitem [{\citenamefont {Yu}\ \emph {et~al.}(2024)\citenamefont {Yu},
  \citenamefont {Suh}, \citenamefont {Roig},\ and\ \citenamefont
  {Agterberg}}]{Yu24}%
  \BibitemOpen
  \bibfield  {author} {\bibinfo {author} {\bibfnamefont {Y.}~\bibnamefont
  {Yu}}, \bibinfo {author} {\bibfnamefont {H.-G.}\ \bibnamefont {Suh}},
  \bibinfo {author} {\bibfnamefont {M.}~\bibnamefont {Roig}}, \ and\ \bibinfo
  {author} {\bibfnamefont {D.~F.}\ \bibnamefont {Agterberg}},\ }\href
  {https://arxiv.org/abs/2402.05180} {\bibfield  {journal} {\bibinfo  {journal}
  {arXiv:2402.05180}\ } (\bibinfo {year} {2024})}\BibitemShut {NoStop}%
\bibitem [{\citenamefont {Wu}\ \emph {et~al.}(2007)\citenamefont {Wu},
  \citenamefont {Sun}, \citenamefont {Fradkin},\ and\ \citenamefont
  {Zhang}}]{Wu07}%
  \BibitemOpen
  \bibfield  {author} {\bibinfo {author} {\bibfnamefont {C.}~\bibnamefont
  {Wu}}, \bibinfo {author} {\bibfnamefont {K.}~\bibnamefont {Sun}}, \bibinfo
  {author} {\bibfnamefont {E.}~\bibnamefont {Fradkin}}, \ and\ \bibinfo
  {author} {\bibfnamefont {S.-C.}\ \bibnamefont {Zhang}},\ }\href {\doibase
  10.1103/PhysRevB.75.115103} {\bibfield  {journal} {\bibinfo  {journal} {Phys.
  Rev. B}\ }\textbf {\bibinfo {volume} {75}},\ \bibinfo {pages} {115103}
  (\bibinfo {year} {2007})}\BibitemShut {NoStop}%
\bibitem [{\citenamefont {{Mazin}}(2022)}]{Mazin22}%
  \BibitemOpen
  \bibfield  {author} {\bibinfo {author} {\bibfnamefont {I.~I.}\ \bibnamefont
  {{Mazin}}},\ }\href {https://arxiv.org/abs/2203.05000} {\bibfield  {journal}
  {\bibinfo  {journal} {arXiv:2203.05000}\ } (\bibinfo {year}
  {2022})}\BibitemShut {NoStop}%
\bibitem [{\citenamefont {Soto-Garrido}\ and\ \citenamefont
  {Fradkin}(2014)}]{Rodrigo14}%
  \BibitemOpen
  \bibfield  {author} {\bibinfo {author} {\bibfnamefont {R.}~\bibnamefont
  {Soto-Garrido}}\ and\ \bibinfo {author} {\bibfnamefont {E.}~\bibnamefont
  {Fradkin}},\ }\href {\doibase 10.1103/PhysRevB.89.165126} {\bibfield
  {journal} {\bibinfo  {journal} {Phys. Rev. B}\ }\textbf {\bibinfo {volume}
  {89}},\ \bibinfo {pages} {165126} (\bibinfo {year} {2014})}\BibitemShut
  {NoStop}%
\bibitem [{\citenamefont {Chakraborty}\ and\ \citenamefont
  {Black-Schaffer}(2024)}]{Chakraborty24}%
  \BibitemOpen
  \bibfield  {author} {\bibinfo {author} {\bibfnamefont {D.}~\bibnamefont
  {Chakraborty}}\ and\ \bibinfo {author} {\bibfnamefont {A.~M.}\ \bibnamefont
  {Black-Schaffer}},\ }\href {https://arxiv.org/abs/2309.14427} {\bibfield
  {journal} {\bibinfo  {journal} {arXiv:2309.14427}\ } (\bibinfo {year}
  {2024})}\BibitemShut {NoStop}%
\bibitem [{\citenamefont {Sumita}\ \emph {et~al.}(2023)\citenamefont {Sumita},
  \citenamefont {Naka},\ and\ \citenamefont {Seo}}]{Shuntaro23}%
  \BibitemOpen
  \bibfield  {author} {\bibinfo {author} {\bibfnamefont {S.}~\bibnamefont
  {Sumita}}, \bibinfo {author} {\bibfnamefont {M.}~\bibnamefont {Naka}}, \ and\
  \bibinfo {author} {\bibfnamefont {H.}~\bibnamefont {Seo}},\ }\href {\doibase
  10.1103/PhysRevResearch.5.043171} {\bibfield  {journal} {\bibinfo  {journal}
  {Phys. Rev. Res.}\ }\textbf {\bibinfo {volume} {5}},\ \bibinfo {pages}
  {043171} (\bibinfo {year} {2023})}\BibitemShut {NoStop}%
\bibitem [{\citenamefont {Bose}\ \emph {et~al.}(2024)\citenamefont {Bose},
  \citenamefont {Vadnais},\ and\ \citenamefont {Paramekanti}}]{bose24}%
  \BibitemOpen
  \bibfield  {author} {\bibinfo {author} {\bibfnamefont {A.}~\bibnamefont
  {Bose}}, \bibinfo {author} {\bibfnamefont {S.}~\bibnamefont {Vadnais}}, \
  and\ \bibinfo {author} {\bibfnamefont {A.}~\bibnamefont {Paramekanti}},\
  }\href {https://arxiv.org/abs/2403.17050} {\bibfield  {journal} {\bibinfo
  {journal} {arXiv:2403.17050}\ } (\bibinfo {year} {2024})}\BibitemShut
  {NoStop}%
\bibitem [{\citenamefont {Sim}\ and\ \citenamefont {Knolle}(2024)}]{sim24}%
  \BibitemOpen
  \bibfield  {author} {\bibinfo {author} {\bibfnamefont {G.}~\bibnamefont
  {Sim}}\ and\ \bibinfo {author} {\bibfnamefont {J.}~\bibnamefont {Knolle}},\
  }\href {https://arxiv.org/abs/2407.01513} {\bibfield  {journal} {\bibinfo
  {journal} {arXiv:2407.01513}\ } (\bibinfo {year} {2024})}\BibitemShut
  {NoStop}%
\bibitem [{\citenamefont {Hong}\ \emph {et~al.}(2024)\citenamefont {Hong},
  \citenamefont {Park},\ and\ \citenamefont {Kim}}]{hong24}%
  \BibitemOpen
  \bibfield  {author} {\bibinfo {author} {\bibfnamefont {S.}~\bibnamefont
  {Hong}}, \bibinfo {author} {\bibfnamefont {M.~J.}\ \bibnamefont {Park}}, \
  and\ \bibinfo {author} {\bibfnamefont {K.-M.}\ \bibnamefont {Kim}},\ }\href
  {https://arxiv.org/abs/2407.02059} {\bibfield  {journal} {\bibinfo  {journal}
  {arXiv:2407.02059}\ } (\bibinfo {year} {2024})}\BibitemShut {NoStop}%
\bibitem [{\citenamefont {Uchida}\ \emph {et~al.}(2020)\citenamefont {Uchida},
  \citenamefont {Nomoto}, \citenamefont {Musashi}, \citenamefont {Arita},\ and\
  \citenamefont {Kawasaki}}]{Uchida20}%
  \BibitemOpen
  \bibfield  {author} {\bibinfo {author} {\bibfnamefont {M.}~\bibnamefont
  {Uchida}}, \bibinfo {author} {\bibfnamefont {T.}~\bibnamefont {Nomoto}},
  \bibinfo {author} {\bibfnamefont {M.}~\bibnamefont {Musashi}}, \bibinfo
  {author} {\bibfnamefont {R.}~\bibnamefont {Arita}}, \ and\ \bibinfo {author}
  {\bibfnamefont {M.}~\bibnamefont {Kawasaki}},\ }\href {\doibase
  10.1103/PhysRevLett.125.147001} {\bibfield  {journal} {\bibinfo  {journal}
  {Phys. Rev. Lett.}\ }\textbf {\bibinfo {volume} {125}},\ \bibinfo {pages}
  {147001} (\bibinfo {year} {2020})}\BibitemShut {NoStop}%
\bibitem [{\citenamefont {\ifmmode~\check{S}\else \v{S}\fi{}mejkal}\ \emph
  {et~al.}(2023)\citenamefont {\ifmmode~\check{S}\else \v{S}\fi{}mejkal},
  \citenamefont {Marmodoro}, \citenamefont {Ahn}, \citenamefont
  {Gonz\'alez-Hern\'andez}, \citenamefont {Turek}, \citenamefont {Mankovsky},
  \citenamefont {Ebert}, \citenamefont {D'Souza}, \citenamefont
  {\ifmmode~\check{S}\else \v{S}\fi{}ipr}, \citenamefont {Sinova},\ and\
  \citenamefont {Jungwirth}}]{Smejkal23}%
  \BibitemOpen
  \bibfield  {author} {\bibinfo {author} {\bibfnamefont {L.}~\bibnamefont
  {\ifmmode~\check{S}\else \v{S}\fi{}mejkal}}, \bibinfo {author} {\bibfnamefont
  {A.}~\bibnamefont {Marmodoro}}, \bibinfo {author} {\bibfnamefont {K.-H.}\
  \bibnamefont {Ahn}}, \bibinfo {author} {\bibfnamefont {R.}~\bibnamefont
  {Gonz\'alez-Hern\'andez}}, \bibinfo {author} {\bibfnamefont {I.}~\bibnamefont
  {Turek}}, \bibinfo {author} {\bibfnamefont {S.}~\bibnamefont {Mankovsky}},
  \bibinfo {author} {\bibfnamefont {H.}~\bibnamefont {Ebert}}, \bibinfo
  {author} {\bibfnamefont {S.~W.}\ \bibnamefont {D'Souza}}, \bibinfo {author}
  {\bibfnamefont {O.~c.~v.}\ \bibnamefont {\ifmmode~\check{S}\else
  \v{S}\fi{}ipr}}, \bibinfo {author} {\bibfnamefont {J.}~\bibnamefont
  {Sinova}}, \ and\ \bibinfo {author} {\bibfnamefont {T.~c.~v.}\ \bibnamefont
  {Jungwirth}},\ }\href {\doibase 10.1103/PhysRevLett.131.256703} {\bibfield
  {journal} {\bibinfo  {journal} {Phys. Rev. Lett.}\ }\textbf {\bibinfo
  {volume} {131}},\ \bibinfo {pages} {256703} (\bibinfo {year}
  {2023})}\BibitemShut {NoStop}%
\bibitem [{\citenamefont {Mazin}\ \emph {et~al.}(2023)\citenamefont {Mazin},
  \citenamefont {González-Hernández},\ and\ \citenamefont
  {Šmejkal}}]{mazin23}%
  \BibitemOpen
  \bibfield  {author} {\bibinfo {author} {\bibfnamefont {I.}~\bibnamefont
  {Mazin}}, \bibinfo {author} {\bibfnamefont {R.}~\bibnamefont
  {González-Hernández}}, \ and\ \bibinfo {author} {\bibfnamefont
  {L.}~\bibnamefont {Šmejkal}},\ }\href {https://arxiv.org/abs/2309.02355}
  {\bibfield  {journal} {\bibinfo  {journal} {arXiv:2309.02355}\ } (\bibinfo
  {year} {2023})}\BibitemShut {NoStop}%
\bibitem [{\citenamefont {Zhang}\ \emph {et~al.}(2024)\citenamefont {Zhang},
  \citenamefont {Hu},\ and\ \citenamefont {Neupert}}]{Zhang24}%
  \BibitemOpen
  \bibfield  {author} {\bibinfo {author} {\bibfnamefont {S.-B.}\ \bibnamefont
  {Zhang}}, \bibinfo {author} {\bibfnamefont {L.-H.}\ \bibnamefont {Hu}}, \
  and\ \bibinfo {author} {\bibfnamefont {T.}~\bibnamefont {Neupert}},\ }\href
  {\doibase 10.1038/s41467-024-45951-3} {\bibfield  {journal} {\bibinfo
  {journal} {Nat. Commun.}\ }\textbf {\bibinfo {volume} {15}},\ \bibinfo
  {pages} {1801} (\bibinfo {year} {2024})}\BibitemShut {NoStop}%
\bibitem [{\citenamefont {Ouassou}\ \emph {et~al.}(2023)\citenamefont
  {Ouassou}, \citenamefont {Brataas},\ and\ \citenamefont
  {Linder}}]{Ouassou23}%
  \BibitemOpen
  \bibfield  {author} {\bibinfo {author} {\bibfnamefont {J.~A.}\ \bibnamefont
  {Ouassou}}, \bibinfo {author} {\bibfnamefont {A.}~\bibnamefont {Brataas}}, \
  and\ \bibinfo {author} {\bibfnamefont {J.}~\bibnamefont {Linder}},\ }\href
  {\doibase 10.1103/PhysRevLett.131.076003} {\bibfield  {journal} {\bibinfo
  {journal} {Phys. Rev. Lett.}\ }\textbf {\bibinfo {volume} {131}},\ \bibinfo
  {pages} {076003} (\bibinfo {year} {2023})}\BibitemShut {NoStop}%
\bibitem [{\citenamefont {Papaj}(2023)}]{Papaj23}%
  \BibitemOpen
  \bibfield  {author} {\bibinfo {author} {\bibfnamefont {M.}~\bibnamefont
  {Papaj}},\ }\href {\doibase 10.1103/PhysRevB.108.L060508} {\bibfield
  {journal} {\bibinfo  {journal} {Phys. Rev. B}\ }\textbf {\bibinfo {volume}
  {108}},\ \bibinfo {pages} {L060508} (\bibinfo {year} {2023})}\BibitemShut
  {NoStop}%
\bibitem [{\citenamefont {Sun}\ \emph {et~al.}(2024)\citenamefont {Sun},
  \citenamefont {Zhang}, \citenamefont {Li},\ and\ \citenamefont
  {Trauzettel}}]{sun24}%
  \BibitemOpen
  \bibfield  {author} {\bibinfo {author} {\bibfnamefont {H.-P.}\ \bibnamefont
  {Sun}}, \bibinfo {author} {\bibfnamefont {S.-B.}\ \bibnamefont {Zhang}},
  \bibinfo {author} {\bibfnamefont {C.-A.}\ \bibnamefont {Li}}, \ and\ \bibinfo
  {author} {\bibfnamefont {B.}~\bibnamefont {Trauzettel}},\ }\href
  {https://arxiv.org/abs/2407.19413} {\bibfield  {journal} {\bibinfo  {journal}
  {arXiv:2407.19413}\ } (\bibinfo {year} {2024})}\BibitemShut {NoStop}%
\bibitem [{\citenamefont {Lu}\ \emph {et~al.}(2024)\citenamefont {Lu},
  \citenamefont {Maeda}, \citenamefont {Ito}, \citenamefont {Yada},\ and\
  \citenamefont {Tanaka}}]{Lu24}%
  \BibitemOpen
  \bibfield  {author} {\bibinfo {author} {\bibfnamefont {B.}~\bibnamefont
  {Lu}}, \bibinfo {author} {\bibfnamefont {K.}~\bibnamefont {Maeda}}, \bibinfo
  {author} {\bibfnamefont {H.}~\bibnamefont {Ito}}, \bibinfo {author}
  {\bibfnamefont {K.}~\bibnamefont {Yada}}, \ and\ \bibinfo {author}
  {\bibfnamefont {Y.}~\bibnamefont {Tanaka}},\ }\href
  {https://arxiv.org/abs/2405.10656} {\bibfield  {journal} {\bibinfo  {journal}
  {arXiv:2405.10656}\ } (\bibinfo {year} {2024})}\BibitemShut {NoStop}%
\bibitem [{\citenamefont {Beenakker}\ and\ \citenamefont
  {Vakhtel}(2023)}]{Beenakker23}%
  \BibitemOpen
  \bibfield  {author} {\bibinfo {author} {\bibfnamefont {C.~W.~J.}\
  \bibnamefont {Beenakker}}\ and\ \bibinfo {author} {\bibfnamefont
  {T.}~\bibnamefont {Vakhtel}},\ }\href {\doibase 10.1103/PhysRevB.108.075425}
  {\bibfield  {journal} {\bibinfo  {journal} {Phys. Rev. B}\ }\textbf {\bibinfo
  {volume} {108}},\ \bibinfo {pages} {075425} (\bibinfo {year}
  {2023})}\BibitemShut {NoStop}%
\bibitem [{\citenamefont {Giil}\ and\ \citenamefont {Linder}(2024)}]{Hans24}%
  \BibitemOpen
  \bibfield  {author} {\bibinfo {author} {\bibfnamefont {H.~G.}\ \bibnamefont
  {Giil}}\ and\ \bibinfo {author} {\bibfnamefont {J.}~\bibnamefont {Linder}},\
  }\href {\doibase 10.1103/PhysRevB.109.134511} {\bibfield  {journal} {\bibinfo
   {journal} {Phys. Rev. B}\ }\textbf {\bibinfo {volume} {109}},\ \bibinfo
  {pages} {134511} (\bibinfo {year} {2024})}\BibitemShut {NoStop}%
\bibitem [{\citenamefont {Banerjee}\ and\ \citenamefont
  {Scheurer}(2024)}]{Banerjee24}%
  \BibitemOpen
  \bibfield  {author} {\bibinfo {author} {\bibfnamefont {S.}~\bibnamefont
  {Banerjee}}\ and\ \bibinfo {author} {\bibfnamefont {M.~S.}\ \bibnamefont
  {Scheurer}},\ }\href {\doibase 10.1103/PhysRevB.110.024503} {\bibfield
  {journal} {\bibinfo  {journal} {Phys. Rev. B}\ }\textbf {\bibinfo {volume}
  {110}},\ \bibinfo {pages} {024503} (\bibinfo {year} {2024})}\BibitemShut
  {NoStop}%
\bibitem [{\citenamefont {Giil}\ \emph {et~al.}(2024)\citenamefont {Giil},
  \citenamefont {Brekke}, \citenamefont {Linder},\ and\ \citenamefont
  {Brataas}}]{giil24}%
  \BibitemOpen
  \bibfield  {author} {\bibinfo {author} {\bibfnamefont {H.~G.}\ \bibnamefont
  {Giil}}, \bibinfo {author} {\bibfnamefont {B.}~\bibnamefont {Brekke}},
  \bibinfo {author} {\bibfnamefont {J.}~\bibnamefont {Linder}}, \ and\ \bibinfo
  {author} {\bibfnamefont {A.}~\bibnamefont {Brataas}},\ }\href
  {https://arxiv.org/abs/2403.04851} {\bibfield  {journal} {\bibinfo  {journal}
  {arXiv:2403.04851}\ } (\bibinfo {year} {2024})}\BibitemShut {NoStop}%
\bibitem [{\citenamefont {Zyuzin}(2024)}]{Zyuzin24}%
  \BibitemOpen
  \bibfield  {author} {\bibinfo {author} {\bibfnamefont {A.~A.}\ \bibnamefont
  {Zyuzin}},\ }\href {\doibase 10.1103/PhysRevB.109.L220505} {\bibfield
  {journal} {\bibinfo  {journal} {Phys. Rev. B}\ }\textbf {\bibinfo {volume}
  {109}},\ \bibinfo {pages} {L220505} (\bibinfo {year} {2024})}\BibitemShut
  {NoStop}%
\bibitem [{\citenamefont {Li}(2024)}]{Li24}%
  \BibitemOpen
  \bibfield  {author} {\bibinfo {author} {\bibfnamefont {Y.-X.}\ \bibnamefont
  {Li}},\ }\href {\doibase 10.1103/PhysRevB.109.224502} {\bibfield  {journal}
  {\bibinfo  {journal} {Phys. Rev. B}\ }\textbf {\bibinfo {volume} {109}},\
  \bibinfo {pages} {224502} (\bibinfo {year} {2024})}\BibitemShut {NoStop}%
\bibitem [{\citenamefont {Ghorashi}\ \emph {et~al.}(2023)\citenamefont
  {Ghorashi}, \citenamefont {Hughes},\ and\ \citenamefont {Cano}}]{Ghorashi23}%
  \BibitemOpen
  \bibfield  {author} {\bibinfo {author} {\bibfnamefont {S.~A.~A.}\
  \bibnamefont {Ghorashi}}, \bibinfo {author} {\bibfnamefont {T.~L.}\
  \bibnamefont {Hughes}}, \ and\ \bibinfo {author} {\bibfnamefont
  {J.}~\bibnamefont {Cano}},\ }\href {https://arxiv.org/abs/2306.09413}
  {\bibfield  {journal} {\bibinfo  {journal} {arXiv:2306.09413}\ } (\bibinfo
  {year} {2023})}\BibitemShut {NoStop}%
\bibitem [{\citenamefont {Zhu}\ \emph {et~al.}(2023)\citenamefont {Zhu},
  \citenamefont {Zhuang}, \citenamefont {Wu},\ and\ \citenamefont
  {Yan}}]{Zhu23}%
  \BibitemOpen
  \bibfield  {author} {\bibinfo {author} {\bibfnamefont {D.}~\bibnamefont
  {Zhu}}, \bibinfo {author} {\bibfnamefont {Z.-Y.}\ \bibnamefont {Zhuang}},
  \bibinfo {author} {\bibfnamefont {Z.}~\bibnamefont {Wu}}, \ and\ \bibinfo
  {author} {\bibfnamefont {Z.}~\bibnamefont {Yan}},\ }\href {\doibase
  10.1103/PhysRevB.108.184505} {\bibfield  {journal} {\bibinfo  {journal}
  {Phys. Rev. B}\ }\textbf {\bibinfo {volume} {108}},\ \bibinfo {pages}
  {184505} (\bibinfo {year} {2023})}\BibitemShut {NoStop}%
\bibitem [{\citenamefont {Nagae}\ \emph {et~al.}(2024)\citenamefont {Nagae},
  \citenamefont {Schnyder},\ and\ \citenamefont {Ikegaya}}]{nagae24}%
  \BibitemOpen
  \bibfield  {author} {\bibinfo {author} {\bibfnamefont {Y.}~\bibnamefont
  {Nagae}}, \bibinfo {author} {\bibfnamefont {A.~P.}\ \bibnamefont {Schnyder}},
  \ and\ \bibinfo {author} {\bibfnamefont {S.}~\bibnamefont {Ikegaya}},\ }\href
  {https://arxiv.org/abs/2403.07117} {\bibfield  {journal} {\bibinfo  {journal}
  {arXiv:2403.07117}\ } (\bibinfo {year} {2024})}\BibitemShut {NoStop}%
\bibitem [{\citenamefont {Sigrist}\ and\ \citenamefont
  {Ueda}(1991)}]{Sigrist91}%
  \BibitemOpen
  \bibfield  {author} {\bibinfo {author} {\bibfnamefont {M.}~\bibnamefont
  {Sigrist}}\ and\ \bibinfo {author} {\bibfnamefont {K.}~\bibnamefont {Ueda}},\
  }\href {\doibase 10.1103/RevModPhys.63.239} {\bibfield  {journal} {\bibinfo
  {journal} {Rev. Mod. Phys.}\ }\textbf {\bibinfo {volume} {63}},\ \bibinfo
  {pages} {239} (\bibinfo {year} {1991})}\BibitemShut {NoStop}%
\bibitem [{\citenamefont {Triola}\ and\ \citenamefont
  {Balatsky}(2016)}]{Triola16}%
  \BibitemOpen
  \bibfield  {author} {\bibinfo {author} {\bibfnamefont {C.}~\bibnamefont
  {Triola}}\ and\ \bibinfo {author} {\bibfnamefont {A.~V.}\ \bibnamefont
  {Balatsky}},\ }\href {\doibase 10.1103/PhysRevB.94.094518} {\bibfield
  {journal} {\bibinfo  {journal} {Phys. Rev. B}\ }\textbf {\bibinfo {volume}
  {94}},\ \bibinfo {pages} {094518} (\bibinfo {year} {2016})}\BibitemShut
  {NoStop}%
\bibitem [{\citenamefont {Linder}\ and\ \citenamefont
  {Balatsky}(2019)}]{Linder19}%
  \BibitemOpen
  \bibfield  {author} {\bibinfo {author} {\bibfnamefont {J.}~\bibnamefont
  {Linder}}\ and\ \bibinfo {author} {\bibfnamefont {A.~V.}\ \bibnamefont
  {Balatsky}},\ }\href {\doibase 10.1103/RevModPhys.91.045005} {\bibfield
  {journal} {\bibinfo  {journal} {Rev. Mod. Phys.}\ }\textbf {\bibinfo {volume}
  {91}},\ \bibinfo {pages} {045005} (\bibinfo {year} {2019})}\BibitemShut
  {NoStop}%
\bibitem [{\citenamefont {Triola}\ \emph {et~al.}(2020)\citenamefont {Triola},
  \citenamefont {Cayao},\ and\ \citenamefont {Black-Schaffer}}]{Triola20}%
  \BibitemOpen
  \bibfield  {author} {\bibinfo {author} {\bibfnamefont {C.}~\bibnamefont
  {Triola}}, \bibinfo {author} {\bibfnamefont {J.}~\bibnamefont {Cayao}}, \
  and\ \bibinfo {author} {\bibfnamefont {A.~M.}\ \bibnamefont
  {Black-Schaffer}},\ }\href {\doibase 10.1002/andp.201900298} {\bibfield
  {journal} {\bibinfo  {journal} {Annalen der Physik}\ }\textbf {\bibinfo
  {volume} {532}},\ \bibinfo {pages} {1900298} (\bibinfo {year}
  {2020})}\BibitemShut {NoStop}%
\bibitem [{Note1()}]{Note1}%
  \BibitemOpen
  \bibinfo {note} {Alternatively, we may treat the sublattice index as an
  orbital degree of freedom \cite {Black-Schaffer13} within an extended unit
  cell consisting of both an A and B site. This results in instead using the
  condition $SPOT =-1$, where $O$ stands for orbital parity, while the spatial
  parity becomes $P=1$ within the unit cell. This results in the same
  conclusions as in Tables \ref {tab:table1}-\ref {tab:table2} but the
  resulting spatial symmetries (form factors) are less straightforward to
  extract.}\BibitemShut {Stop}%
\bibitem [{\citenamefont {Berezinskii}(1974)}]{Berezinskii74}%
  \BibitemOpen
  \bibfield  {author} {\bibinfo {author} {\bibfnamefont {V.~L.}\ \bibnamefont
  {Berezinskii}},\ }\href@noop {} {\bibfield  {journal} {\bibinfo  {journal}
  {Pisma Zh. Eksp. Teor. Fiz. 20, 628}\ }\textbf {\bibinfo {volume} {20}},\
  \bibinfo {pages} {628} (\bibinfo {year} {1974})}\BibitemShut {NoStop}%
\bibitem [{\citenamefont {Kirkpatrick}\ and\ \citenamefont
  {Belitz}(1991)}]{Kirkpatrick91}%
  \BibitemOpen
  \bibfield  {author} {\bibinfo {author} {\bibfnamefont {T.~R.}\ \bibnamefont
  {Kirkpatrick}}\ and\ \bibinfo {author} {\bibfnamefont {D.}~\bibnamefont
  {Belitz}},\ }\href {\doibase 10.1103/PhysRevLett.66.1533} {\bibfield
  {journal} {\bibinfo  {journal} {Phys. Rev. Lett.}\ }\textbf {\bibinfo
  {volume} {66}},\ \bibinfo {pages} {1533} (\bibinfo {year}
  {1991})}\BibitemShut {NoStop}%
\bibitem [{\citenamefont {Balatsky}\ and\ \citenamefont
  {Abrahams}(1992)}]{Balatsky92}%
  \BibitemOpen
  \bibfield  {author} {\bibinfo {author} {\bibfnamefont {A.}~\bibnamefont
  {Balatsky}}\ and\ \bibinfo {author} {\bibfnamefont {E.}~\bibnamefont
  {Abrahams}},\ }\href {\doibase 10.1103/PhysRevB.45.13125} {\bibfield
  {journal} {\bibinfo  {journal} {Phys. Rev. B}\ }\textbf {\bibinfo {volume}
  {45}},\ \bibinfo {pages} {13125} (\bibinfo {year} {1992})}\BibitemShut
  {NoStop}%
\bibitem [{\citenamefont {Schrieffer}\ \emph {et~al.}(1994)\citenamefont
  {Schrieffer}, \citenamefont {Balatsky}, \citenamefont {Abrahams},\ and\
  \citenamefont {Scalapino}}]{Schrieffer94}%
  \BibitemOpen
  \bibfield  {author} {\bibinfo {author} {\bibfnamefont {J.~R.}\ \bibnamefont
  {Schrieffer}}, \bibinfo {author} {\bibfnamefont {A.~V.}\ \bibnamefont
  {Balatsky}}, \bibinfo {author} {\bibfnamefont {E.}~\bibnamefont {Abrahams}},
  \ and\ \bibinfo {author} {\bibfnamefont {D.~J.}\ \bibnamefont {Scalapino}},\
  }\href {\doibase 10.1007/BF00728448} {\bibfield  {journal} {\bibinfo
  {journal} {J. Supercond.}\ }\textbf {\bibinfo {volume} {7}},\ \bibinfo
  {pages} {501} (\bibinfo {year} {1994})}\BibitemShut {NoStop}%
\bibitem [{\citenamefont {Bergeret}\ \emph {et~al.}(2005)\citenamefont
  {Bergeret}, \citenamefont {Volkov},\ and\ \citenamefont
  {Efetov}}]{Bergeret05}%
  \BibitemOpen
  \bibfield  {author} {\bibinfo {author} {\bibfnamefont {F.~S.}\ \bibnamefont
  {Bergeret}}, \bibinfo {author} {\bibfnamefont {A.~F.}\ \bibnamefont
  {Volkov}}, \ and\ \bibinfo {author} {\bibfnamefont {K.~B.}\ \bibnamefont
  {Efetov}},\ }\href {\doibase 10.1103/RevModPhys.77.1321} {\bibfield
  {journal} {\bibinfo  {journal} {Rev. Mod. Phys.}\ }\textbf {\bibinfo {volume}
  {77}},\ \bibinfo {pages} {1321} (\bibinfo {year} {2005})}\BibitemShut
  {NoStop}%
\bibitem [{\citenamefont {Yokoyama}\ \emph {et~al.}(2011)\citenamefont
  {Yokoyama}, \citenamefont {Tanaka},\ and\ \citenamefont
  {Nagaosa}}]{Yokoyama11}%
  \BibitemOpen
  \bibfield  {author} {\bibinfo {author} {\bibfnamefont {T.}~\bibnamefont
  {Yokoyama}}, \bibinfo {author} {\bibfnamefont {Y.}~\bibnamefont {Tanaka}}, \
  and\ \bibinfo {author} {\bibfnamefont {N.}~\bibnamefont {Nagaosa}},\ }\href
  {\doibase 10.1103/PhysRevLett.106.246601} {\bibfield  {journal} {\bibinfo
  {journal} {Phys. Rev. Lett.}\ }\textbf {\bibinfo {volume} {106}},\ \bibinfo
  {pages} {246601} (\bibinfo {year} {2011})}\BibitemShut {NoStop}%
\bibitem [{\citenamefont {Black-Schaffer}\ and\ \citenamefont
  {Balatsky}(2013)}]{Black-Schaffer13}%
  \BibitemOpen
  \bibfield  {author} {\bibinfo {author} {\bibfnamefont {A.~M.}\ \bibnamefont
  {Black-Schaffer}}\ and\ \bibinfo {author} {\bibfnamefont {A.~V.}\
  \bibnamefont {Balatsky}},\ }\href {\doibase 10.1103/PhysRevB.88.104514}
  {\bibfield  {journal} {\bibinfo  {journal} {Phys. Rev. B}\ }\textbf {\bibinfo
  {volume} {88}},\ \bibinfo {pages} {104514} (\bibinfo {year}
  {2013})}\BibitemShut {NoStop}%
\bibitem [{\citenamefont {Alidoust}\ \emph {et~al.}(2014)\citenamefont
  {Alidoust}, \citenamefont {Halterman},\ and\ \citenamefont
  {Linder}}]{Alidoust14}%
  \BibitemOpen
  \bibfield  {author} {\bibinfo {author} {\bibfnamefont {M.}~\bibnamefont
  {Alidoust}}, \bibinfo {author} {\bibfnamefont {K.}~\bibnamefont {Halterman}},
  \ and\ \bibinfo {author} {\bibfnamefont {J.}~\bibnamefont {Linder}},\ }\href
  {\doibase 10.1103/PhysRevB.89.054508} {\bibfield  {journal} {\bibinfo
  {journal} {Phys. Rev. B}\ }\textbf {\bibinfo {volume} {89}},\ \bibinfo
  {pages} {054508} (\bibinfo {year} {2014})}\BibitemShut {NoStop}%
\bibitem [{\citenamefont {Tanaka}\ \emph {et~al.}(2012)\citenamefont {Tanaka},
  \citenamefont {Sato},\ and\ \citenamefont {Nagaosa}}]{Tanaka12}%
  \BibitemOpen
  \bibfield  {author} {\bibinfo {author} {\bibfnamefont {Y.}~\bibnamefont
  {Tanaka}}, \bibinfo {author} {\bibfnamefont {M.}~\bibnamefont {Sato}}, \ and\
  \bibinfo {author} {\bibfnamefont {N.}~\bibnamefont {Nagaosa}},\ }\href
  {\doibase 10.1143/JPSJ.81.011013} {\bibfield  {journal} {\bibinfo  {journal}
  {J. Phys. Soc. Jpn.}\ }\textbf {\bibinfo {volume} {81}},\ \bibinfo {pages}
  {011013} (\bibinfo {year} {2012})}\BibitemShut {NoStop}%
\bibitem [{\citenamefont {Chakraborty}\ and\ \citenamefont
  {Black-Schaffer}(2021)}]{Chakraborty21}%
  \BibitemOpen
  \bibfield  {author} {\bibinfo {author} {\bibfnamefont {D.}~\bibnamefont
  {Chakraborty}}\ and\ \bibinfo {author} {\bibfnamefont {A.~M.}\ \bibnamefont
  {Black-Schaffer}},\ }\href {\doibase 10.1088/1367-2630/abe15d} {\bibfield
  {journal} {\bibinfo  {journal} {New J. Phys.}\ }\textbf {\bibinfo {volume}
  {23}},\ \bibinfo {pages} {033001} (\bibinfo {year} {2021})}\BibitemShut
  {NoStop}%
\bibitem [{\citenamefont {Chakraborty}\ and\ \citenamefont
  {Black-Schaffer}(2022{\natexlab{a}})}]{Chakraborty22a}%
  \BibitemOpen
  \bibfield  {author} {\bibinfo {author} {\bibfnamefont {D.}~\bibnamefont
  {Chakraborty}}\ and\ \bibinfo {author} {\bibfnamefont {A.~M.}\ \bibnamefont
  {Black-Schaffer}},\ }\href {\doibase 10.1103/PhysRevLett.129.247001}
  {\bibfield  {journal} {\bibinfo  {journal} {Phys. Rev. Lett.}\ }\textbf
  {\bibinfo {volume} {129}},\ \bibinfo {pages} {247001} (\bibinfo {year}
  {2022}{\natexlab{a}})}\BibitemShut {NoStop}%
\bibitem [{\citenamefont {Chakraborty}\ and\ \citenamefont
  {Black-Schaffer}(2022{\natexlab{b}})}]{Chakraborty22b}%
  \BibitemOpen
  \bibfield  {author} {\bibinfo {author} {\bibfnamefont {D.}~\bibnamefont
  {Chakraborty}}\ and\ \bibinfo {author} {\bibfnamefont {A.~M.}\ \bibnamefont
  {Black-Schaffer}},\ }\href {\doibase 10.1103/PhysRevB.106.024511} {\bibfield
  {journal} {\bibinfo  {journal} {Phys. Rev. B}\ }\textbf {\bibinfo {volume}
  {106}},\ \bibinfo {pages} {024511} (\bibinfo {year}
  {2022}{\natexlab{b}})}\BibitemShut {NoStop}%
\bibitem [{\citenamefont {Bergeret}\ \emph
  {et~al.}(2001{\natexlab{a}})\citenamefont {Bergeret}, \citenamefont
  {Volkov},\ and\ \citenamefont {Efetov}}]{Bergeret01}%
  \BibitemOpen
  \bibfield  {author} {\bibinfo {author} {\bibfnamefont {F.~S.}\ \bibnamefont
  {Bergeret}}, \bibinfo {author} {\bibfnamefont {A.~F.}\ \bibnamefont
  {Volkov}}, \ and\ \bibinfo {author} {\bibfnamefont {K.~B.}\ \bibnamefont
  {Efetov}},\ }\href {\doibase 10.1103/PhysRevB.64.134506} {\bibfield
  {journal} {\bibinfo  {journal} {Phys. Rev. B}\ }\textbf {\bibinfo {volume}
  {64}},\ \bibinfo {pages} {134506} (\bibinfo {year}
  {2001}{\natexlab{a}})}\BibitemShut {NoStop}%
\bibitem [{\citenamefont {Bergeret}\ \emph
  {et~al.}(2001{\natexlab{b}})\citenamefont {Bergeret}, \citenamefont
  {Volkov},\ and\ \citenamefont {Efetov}}]{Bergeret01prb}%
  \BibitemOpen
  \bibfield  {author} {\bibinfo {author} {\bibfnamefont {F.~S.}\ \bibnamefont
  {Bergeret}}, \bibinfo {author} {\bibfnamefont {A.~F.}\ \bibnamefont
  {Volkov}}, \ and\ \bibinfo {author} {\bibfnamefont {K.~B.}\ \bibnamefont
  {Efetov}},\ }\href {\doibase 10.1103/PhysRevB.64.134506} {\bibfield
  {journal} {\bibinfo  {journal} {Phys. Rev. B}\ }\textbf {\bibinfo {volume}
  {64}},\ \bibinfo {pages} {134506} (\bibinfo {year}
  {2001}{\natexlab{b}})}\BibitemShut {NoStop}%
\bibitem [{\citenamefont {Buzdin}(2005)}]{Buzdin05}%
  \BibitemOpen
  \bibfield  {author} {\bibinfo {author} {\bibfnamefont {A.~I.}\ \bibnamefont
  {Buzdin}},\ }\href {\doibase 10.1103/RevModPhys.77.935} {\bibfield  {journal}
  {\bibinfo  {journal} {Rev. Mod. Phys.}\ }\textbf {\bibinfo {volume} {77}},\
  \bibinfo {pages} {935} (\bibinfo {year} {2005})}\BibitemShut {NoStop}%
\bibitem [{\citenamefont {Di~Bernardo}\ \emph
  {et~al.}(2015{\natexlab{a}})\citenamefont {Di~Bernardo}, \citenamefont
  {Diesch}, \citenamefont {Gu}, \citenamefont {Linder}, \citenamefont
  {Divitini}, \citenamefont {Ducati}, \citenamefont {Scheer}, \citenamefont
  {Blamire},\ and\ \citenamefont {Robinson}}]{DiBernardo15}%
  \BibitemOpen
  \bibfield  {author} {\bibinfo {author} {\bibfnamefont {A.}~\bibnamefont
  {Di~Bernardo}}, \bibinfo {author} {\bibfnamefont {S.}~\bibnamefont {Diesch}},
  \bibinfo {author} {\bibfnamefont {Y.}~\bibnamefont {Gu}}, \bibinfo {author}
  {\bibfnamefont {J.}~\bibnamefont {Linder}}, \bibinfo {author} {\bibfnamefont
  {G.}~\bibnamefont {Divitini}}, \bibinfo {author} {\bibfnamefont
  {C.}~\bibnamefont {Ducati}}, \bibinfo {author} {\bibfnamefont
  {E.}~\bibnamefont {Scheer}}, \bibinfo {author} {\bibfnamefont {M.~G.}\
  \bibnamefont {Blamire}}, \ and\ \bibinfo {author} {\bibfnamefont {J.~W.~A.}\
  \bibnamefont {Robinson}},\ }\href {\doibase 10.1038/ncomms9053} {\bibfield
  {journal} {\bibinfo  {journal} {Nat. Commun.}\ }\textbf {\bibinfo {volume}
  {6}},\ \bibinfo {pages} {8053} (\bibinfo {year}
  {2015}{\natexlab{a}})}\BibitemShut {NoStop}%
\bibitem [{\citenamefont {Di~Bernardo}\ \emph
  {et~al.}(2015{\natexlab{b}})\citenamefont {Di~Bernardo}, \citenamefont
  {Salman}, \citenamefont {Wang}, \citenamefont {Amado}, \citenamefont
  {Egilmez}, \citenamefont {Flokstra}, \citenamefont {Suter}, \citenamefont
  {Lee}, \citenamefont {Zhao}, \citenamefont {Prokscha}, \citenamefont
  {Morenzoni}, \citenamefont {Blamire}, \citenamefont {Linder},\ and\
  \citenamefont {Robinson}}]{Bernardo15}%
  \BibitemOpen
  \bibfield  {author} {\bibinfo {author} {\bibfnamefont {A.}~\bibnamefont
  {Di~Bernardo}}, \bibinfo {author} {\bibfnamefont {Z.}~\bibnamefont {Salman}},
  \bibinfo {author} {\bibfnamefont {X.~L.}\ \bibnamefont {Wang}}, \bibinfo
  {author} {\bibfnamefont {M.}~\bibnamefont {Amado}}, \bibinfo {author}
  {\bibfnamefont {M.}~\bibnamefont {Egilmez}}, \bibinfo {author} {\bibfnamefont
  {M.~G.}\ \bibnamefont {Flokstra}}, \bibinfo {author} {\bibfnamefont
  {A.}~\bibnamefont {Suter}}, \bibinfo {author} {\bibfnamefont {S.~L.}\
  \bibnamefont {Lee}}, \bibinfo {author} {\bibfnamefont {J.~H.}\ \bibnamefont
  {Zhao}}, \bibinfo {author} {\bibfnamefont {T.}~\bibnamefont {Prokscha}},
  \bibinfo {author} {\bibfnamefont {E.}~\bibnamefont {Morenzoni}}, \bibinfo
  {author} {\bibfnamefont {M.~G.}\ \bibnamefont {Blamire}}, \bibinfo {author}
  {\bibfnamefont {J.}~\bibnamefont {Linder}}, \ and\ \bibinfo {author}
  {\bibfnamefont {J.~W.~A.}\ \bibnamefont {Robinson}},\ }\href {\doibase
  10.1103/PhysRevX.5.041021} {\bibfield  {journal} {\bibinfo  {journal} {Phys.
  Rev. X}\ }\textbf {\bibinfo {volume} {5}},\ \bibinfo {pages} {041021}
  (\bibinfo {year} {2015}{\natexlab{b}})}\BibitemShut {NoStop}%
\bibitem [{\citenamefont {Jacobsen}\ \emph {et~al.}(2016)\citenamefont
  {Jacobsen}, \citenamefont {Kulagina},\ and\ \citenamefont
  {Linder}}]{Jacobsen16}%
  \BibitemOpen
  \bibfield  {author} {\bibinfo {author} {\bibfnamefont {S.~H.}\ \bibnamefont
  {Jacobsen}}, \bibinfo {author} {\bibfnamefont {I.}~\bibnamefont {Kulagina}},
  \ and\ \bibinfo {author} {\bibfnamefont {J.}~\bibnamefont {Linder}},\ }\href
  {\doibase 10.1038/srep23926} {\bibfield  {journal} {\bibinfo  {journal}
  {Scientific Reports}\ }\textbf {\bibinfo {volume} {6}},\ \bibinfo {pages}
  {23926} (\bibinfo {year} {2016})}\BibitemShut {NoStop}%
\bibitem [{\citenamefont {Ouassou}\ \emph {et~al.}(2017)\citenamefont
  {Ouassou}, \citenamefont {Jacobsen},\ and\ \citenamefont
  {Linder}}]{Ouassou17}%
  \BibitemOpen
  \bibfield  {author} {\bibinfo {author} {\bibfnamefont {J.~A.}\ \bibnamefont
  {Ouassou}}, \bibinfo {author} {\bibfnamefont {S.~H.}\ \bibnamefont
  {Jacobsen}}, \ and\ \bibinfo {author} {\bibfnamefont {J.}~\bibnamefont
  {Linder}},\ }\href {\doibase 10.1103/PhysRevB.96.094505} {\bibfield
  {journal} {\bibinfo  {journal} {Phys. Rev. B}\ }\textbf {\bibinfo {volume}
  {96}},\ \bibinfo {pages} {094505} (\bibinfo {year} {2017})}\BibitemShut
  {NoStop}%
\bibitem [{\citenamefont {Perrin}\ \emph {et~al.}(2020)\citenamefont {Perrin},
  \citenamefont {Santos}, \citenamefont {M\'enard}, \citenamefont {Brun},
  \citenamefont {Cren}, \citenamefont {Civelli},\ and\ \citenamefont
  {Simon}}]{Perrin20}%
  \BibitemOpen
  \bibfield  {author} {\bibinfo {author} {\bibfnamefont {V.}~\bibnamefont
  {Perrin}}, \bibinfo {author} {\bibfnamefont {F.~L.~N.}\ \bibnamefont
  {Santos}}, \bibinfo {author} {\bibfnamefont {G.~C.}\ \bibnamefont
  {M\'enard}}, \bibinfo {author} {\bibfnamefont {C.}~\bibnamefont {Brun}},
  \bibinfo {author} {\bibfnamefont {T.}~\bibnamefont {Cren}}, \bibinfo {author}
  {\bibfnamefont {M.}~\bibnamefont {Civelli}}, \ and\ \bibinfo {author}
  {\bibfnamefont {P.}~\bibnamefont {Simon}},\ }\href {\doibase
  10.1103/PhysRevLett.125.117003} {\bibfield  {journal} {\bibinfo  {journal}
  {Phys. Rev. Lett.}\ }\textbf {\bibinfo {volume} {125}},\ \bibinfo {pages}
  {117003} (\bibinfo {year} {2020})}\BibitemShut {NoStop}%
\bibitem [{\citenamefont {Fossheim}\ and\ \citenamefont
  {Sudb\o{}}(2005)}]{SudboBook}%
  \BibitemOpen
  \bibfield  {author} {\bibinfo {author} {\bibfnamefont {K.}~\bibnamefont
  {Fossheim}}\ and\ \bibinfo {author} {\bibfnamefont {A.}~\bibnamefont
  {Sudb\o{}}},\ }\href {https://books.google.se/books?id=Ep1MLS9YQX8C} {\emph
  {\bibinfo {title} {Superconductivity: Physics and Applications}}}\ (\bibinfo
  {publisher} {Wiley},\ \bibinfo {year} {2005})\BibitemShut {NoStop}%
\end{thebibliography}%

\end{document}